\documentclass[a4paper]{article}
\usepackage{graphicx} 
\usepackage{subcaption}
\usepackage{color}
\usepackage{amsfonts}
\usepackage{amsmath}
\usepackage{mathtools}
\usepackage{physics}
\usepackage{dcolumn}
\usepackage{bm}
\usepackage{cancel}
\usepackage{tensor}
\usepackage{amsthm}
\usepackage{sectsty}
\usepackage[toc,page]{appendix}

\newtheorem{lemma}{Lemma}
\newtheorem{proposition}{Proposition}
\newtheorem{remark}{Remark}
\newtheorem{theorem}{Theorem}

\allowdisplaybreaks

\newcounter{mnotecount}

\newcommand{\mnotex}[1]
{\protect{\stepcounter{mnotecount}}$^{\mbox{\footnotesize $\bullet$\themnotecount}}$ 
\marginpar{
\raggedright\tiny\em
$\!\!\!\!\!\!\,\bullet$\themnotecount: #1} }

\def\bma{{\bm a}}
\def\bmb{{\bm b}}
\def\bmc{{\bm c}}
\def\bmd{{\bm d}}
\def\bme{{\bm e}}
\def\bmf{{\bm f}}
\def\bmg{{\bm g}}

\def\bmi{{\bm i}}
\def\bmj{{\bm j}}
\def\bmk{{\bm k}}
\def\bml{{\bm l}}
\def\bmn{{\bm n}}
\def\bmm{{\bm m}}

\def\bmp{{\bm p}}
\def\bmq{{\bm q}}

\def\bmu{{\bm u}}
\def\bmv{{\bm v}}

\def\bmzero{{\bm 0}}

\def\bmomega{{\bm \omega}}

\def\bmA{{\bm A}}
\def\bmB{{\bm B}}
\def\bmC{{\bm C}}
\def\bmD{{\bm D}}
\def\bmE{{\bm E}}

\begin{document}

\title{\textbf{Evolution equations for a wide range of Einstein-matter systems}}

\author{M Normann\footnote{\tt m.normann@qmul.ac.uk} and JA Valiente Kroon\footnote{\tt j.a.valiente-kroon@qmul.ac.uk}\\
{\em School of Mathematical Sciences}\\
{\em Queen Mary, University of London} \\
{\em Mile End Road, London E1 4NS, UK}}


\date{\today}

\maketitle

\begin{abstract}
We use an orthonormal frame approach to provide a general framework for the first order hyperbolic reduction of the Einstein equations coupled to a fairly generic class of matter models. Our analysis covers the special cases of dust and perfect fluid. We also provide a discussion of self-gravitating elastic matter. The frame is Fermi-Walker propagated and coordinates
are chosen such as to satisfy the Lagrange condition. We show the propagation of the constraints of the Einstein-matter system.
\end{abstract}

\section{Introduction}

Einstein's theory of General Relativity provides us with the most appropriate tool for studying the dynamics of self gravitating objects. It is therefore of clear interest to study the structural properties of the Einstein field equations and to provide a framework for studying their solutions. The \emph{Cauchy problem} (or, \emph{initial value problem}) provides a setting for the analysis of generic solutions to the field equations parametrised in terms of the initial conditions ---for details, see \cite{Choquet-Bruhat2014,Friedrich2000,Rendall}. In particular, one is interested in showing that the Einstein equations admit a well-posed initial value formulation \cite{Wald1984}. The standard strategy to address this issue is to show that the Einstein equations imply evolution equations that are on a \emph{hyperbolic form}. Physical considerations associated to causality lead to the expectation of the Einstein equations admitting a hyperbolic formulation despite the fact that the immediate form of the equations is not manifestly hyperbolic due to general covariance. Thus, it is necessary to find a subset of the Einstein equations which indeed admits hyperbolicity. This procedure is called \textit{hyperbolic reduction} ---see \cite{Kroon2016} and \cite{Friedrich1996} for details; for an overview of the different reduction methods, see \cite{Reula1998}. 

The well-posedness of the vacuum Einstein equations was first shown in \cite{Foures-Bruhat1952} ---and later in the case for dust and the Einstein-Euler by the same author \cite{Foures-Bruhat1958}. These results were
obtained using a \textit{harmonic gauge} to reduce the field equations
to a form which is second order hyperbolic (Leray hyperbolicity). In \cite{Andersson2016} this method is extended to show existence of solutions locally for a self-gravitating, relativistic elastic body
with compact support. Furthermore, in \cite{Disconzi2014} well-posedness of a viscous fluid coupled to the Einstein equations is presented and in \cite{Disconzi2015} a viable first order system is constructed. In \cite{Friedrichs1954} the concept of \emph{first order symmetric hyperbolic} (FOSH) equations was developed. The same author showed later \cite{Friedrichs1974} that the Einstein-Euler system could be put on a FOSH form. In \cite{Friedrich1998} a different approach, which
makes use of a formulation in terms of frame fields, is
employed to construct evolution equations for the Einstein-Euler system which also are on the form of a FOSH system. This
method has the advantage that the reduced equations are symmetric
hyperbolic while still maintaining a Lagrangian form ---which is
important in order to keep track of a boundary in the case of matter distributions with compact
support.

As is made clear by the discussion above, it has been customary up until now to apply the hyperbolic reduction procedure to individual matter models separately ---i.e. for every particular type of energy-momentum tensor. The motivation for our study is provided by the observation that the energy momentum tensor for a perfect fluid, elastic matter ---see \cite{Beig2003a,J.Kijowski1992,B.Carter1972} for details--- and bulk viscosity ---e.g. see \cite{Bemfica2019,Normann2016,PhysRevD.91.043532,Brevik1997,Padmanabhan1987} and references therein--- may be put on a form consisting of a part involving the 4-velocity $\bmu$ and energy density $\rho$ and a part involving a spatial symmetric tensor $\bm\Pi$.
 Thus, by "hiding" the specific matter variables in the tensor $\bm\Pi$ one cannot differentiate between elastic matter, perfect or viscous fluid by considering the energy momentum tensor alone. By employing a hyperbolic reduction of the Einstein field equations coupled to an energy-momentum tensor on such a general form, we provide the necessary conditions for such a matter model to form FOSH evolution equations. \textit{ We show that one can avoid the details of the specific matter models in the construction of a FOSH system by introducing an auxilliary field. Furthermore, we provide a constraint on the $\bm\Pi$ on the form of a wave equation, which must be satisfied for the system to be put on a FOSH form}.

The procedure we employ to obtain these evolution equations is similar to that of \cite{Friedrich1998} and may be described as follows: we introduce a frame field to replace the metric tensor as a variable and fix the gauge by choosing Lagrangian coordinates ---i.e. one of the vectors of the frame field is chosen as to coincide with the 4-velocity of the particle trajectories; we also let the rest of the frame be Fermi propagated. By virtue of the Bianchi identity and assuming the connection to be Levi-Civita we show that the solution to a set of new field equations constructed with so called \textit{zero-quantities} implies the existence of a metric solution to the Einstein field equations. A subset of these equations provides the symmetric hyperbolic evolution equations. As part of this construction, it turns out to be necessary to introduce an auxilliary field to remove derivatives of the energy-momentum tensor from the principal part of the evolution equation of some of the geometric fields. The evolution equation of $\bm\Pi$ ---which encodes the matter fields--- is given in terms of the electric decomposition of the auxilliary field. Finally, we make use of the evolution equations, Cartan's identity and the Bianchi identities to show the propagation of constraints. It is important to stress that due to the generality of the procedure, we do not provide an equation defining $\rho$. It is therefore necessary to provide an equation of state (or the equivalent) when using our equations for a specific matter model. We treat dust and perfect fluid as examples at the end and briefly discuss elastic matter.

A limitations of our procedure is in requirement of $\bm\Pi$ being a purely spatial tensor ---indeed, without this requirement the energy momentum tensor would take its most general decomposition form. The difficulty of allowing $\bm\Pi$ to have timelike components resides in the procedure of keeping the hyperbolicity of the theory. We have used the spatial property of $\bm\Pi$ extensively in the process of eliminating problematic derivative terms from the principal part of the equations. We also assume that the equations of motion for a matter system may be entirely determined by the divergence-free condition of the energy-momentum tensor. Thus, any matter models which require additional equations to close the evolution of the matter variables, are not considered herein.

Lastly, we should mention that a very good discussion of the Einstein-Euler-entropy system is found in \cite{Disconzi2013} where a complete discussion of the arguments of the framework put forward in \cite{Friedrich1998} is given.


\subsection*{Overview of the article}
In Section \ref{section1} we introduce the geometric tools necessary for the subsequent discussion; in particular the frame formalism is introduced. The Einstein equations together with the energy momentum tensor is presented in Section \ref{section2}; we also give a brief review of the projection formalism. In Section \ref{section3} we outline the gauge choices and in Section \ref{section4} we present the zero-quantities used in the propagation of constraints. The evolution equations for our system are derived in Section \ref{section5}. In Section \ref{section6} we show propagation of constraints and in Section \ref{section7} we present the reduced equations for the special cases of dust and perfect fluid. A brief discussion of self-gravitating elastic matter is also presented. Final remarks are given in Section \ref{section8}. An appendix provides an extended discussion of a framework for relativistic elasticity ---this model provided the main motivation for the present analysis.

\subsection*{Notation and conventions}
Throughout, for covenience, we use a combination of abstract-index and index-free notation to denote the various tensorial objects. Greek and Latin letters will be used as coordinate
indices in the spacetime manifold, where $\mu,\,\nu,\,\lambda,\ldots =
\{0,1,2,3\}$ and $i,j,k \ldots = \{1,2,3\}$. To denote frame indices
we will make use of bold latin letters where $\bma, \bmb, \bmc \ldots
= \{0,1,2,3\}$ and $\bmi, \bmj, \bmk \ldots = \{1,2,3\}$. Hence, the
components in a frame basis of a vector $\bmv \in \mathcal{M}$ is thus
labelled $v^{\mathbf{a}}$.

\section{Geometric background}
\label{section1}
In what follows, let $(\mathcal{M},\bmg)$ denote a spacetime represented by a 4-dimensional
manifold, $\mathcal{M}$, with a Lorentzian metric $\bmg$. The
motion of particles of some matter filling spacetime give rise to a
natural splitting by constructing frames comoving with the flow lines
of the particles. One advantage with such a view is that it does not
require a foliation. We shall denote the tangent vector to the flow
lines as $\bmu$ satisfying
\[
\bmg(\bmu, \bmu) = -1.
\]
At each point $p \in \mathcal{M}$ the frame field
$\{\bme_{\bma}\}$ is such that
\[
\bmg(\bme_{\bma},\bme_{\bmb}) = \eta_{\bma\bmb}.
\]
The frames $\{\bme_{\bma}\}$ give rise to a co-frame, $\{\mathbf{\omega}^{\bma}\}$ satisfying
\[
\langle{\bme_{\bma}, \mathbf{\bmomega}^{\bmb}\rangle} = \delta_{\bma}{}^{\bmb}.
\]
In the following all indices will be given in terms of the frame and co-frame unless otherwise stated. The metric tensor give rise to a natural connection $\mathbf{\nabla}$ such that $\mathbf{\nabla} \bmg = 0$,
which is the \textit{metric compatibility condition}. In terms of the frames, this condition takes the form
\begin{equation}
\label{metricComp}
\Gamma_{\bma}{}^{\bmb}{}_{\bmc} \eta_{\bmb\bmd} + \Gamma_{\bma}{}^{\bmb}{}_{\bmd} \eta_{\bmb\bmc} = 0,
\end{equation}
where the \textit{frame connection coefficients} are defined by the directional derivative along the direction of the frame indices
\[
\nabla_{\bma} \bme_{\bmb} = \Gamma_{\bma}{}^{\bmc}{}_{\bmb} \bme_{\bmc}, \qquad \nabla_{\bma} = \langle{\bme_{\bma}, \mathbf{\nabla}\rangle}.
\]
Furthermore, if the connection $\mathbf{\nabla}$ is \textit{torsion-free}, we have that
\begin{equation}
\label{torsionFree}
\Sigma_{\bma}{}^{\bmc}{}_{\bmb} = 0,
\end{equation}
where the frame components of the \textit{torsion tensor} are defined by
\[
\Sigma_{\bma}{}^{\bmc}{}_{\bmb} \bme_{\bmc} = \left[\bme_{\bma}, \bme_{\bmb}\right] + \left(\Gamma_{\bma}{}^{\bmc}{}_{\bmb} - \Gamma_{\bmb}{}^{\bmc}{}_{\bma}\right) \bme_{\bmc}.
\]
The commutation of the connection may be expressed in terms of the \textit{Riemann curvature tensor} and the torsion tensor
\begin{eqnarray*}
&&\nabla_{[\bma}\nabla_{\bmb]}v^{\bmc} = R^{\bmc}{}_{\bmd\bma\bmb} v^{\bmd} +\Sigma_{\bma}{}^{\bmd}{}_{\bmb}\nabla_{\bmd}v^{\bmc},\\
&&\nabla_{[\bma}\nabla_{\bmb]}w_{\bmc} = -R^{\bmd}{}_{\bmc\bma\bmb} w_{\bmd} +\Sigma_{\bma}{}^{\bmd}{}_{\bmb}\nabla_{\bmd} w_{\bmc}\label{CommutatorRiemann}.
\end{eqnarray*}
The frame components of the Riemann curvature tensor is given by
\begin{equation}
R^\bmc{}_{\bmd\bma\bmb} = \partial_\bma\Gamma_\bmb{}^\bmc{}_\bmd
   - \partial_\bmb \Gamma_\bma{}^\bmc{}_\bmd +
   \Gamma_\bmf{}^\bmc{}_\bmd(\Gamma_\bmb{}^\bmf{}_\bma -
   \Gamma_\bma{}^\bmf{}_\bmb) +
   \Gamma_\bmb{}^\bmf{}_\bmd\Gamma_\bma{}^\bmc{}_\bmf -
   \Gamma_\bma{}^\bmf{}_\bmd\Gamma_\bmb{}^\bmc{}_\bmf
   -\Sigma_\bma{}^\bmf{}_\bmb \Gamma_\bmf{}^\bmc{}_\bmd \label{RiemannExpansion}
\end{equation}
---see \cite{Kroon2016} for details. The Riemann tensor has all the usual symmetries, and it satisfies the
\textit{Bianchi identity} for a general connection
\begin{eqnarray}
&& R^{\bmd}{}_{[\bmc\bma\bmb]} + \nabla_{[\bma}\Sigma_{\bmb}{}^{\bmd}{}_{\bmc]} +
   \Sigma_{[\bma}{}^{\bme}{}_{\bmb}\Sigma_{\bmc]}{}^{\bmd}{}_{\bme}=0, \label{1stBianchiId}\\
&& \nabla_{[\bma} R^{\bmd}{}_{|\bme|\bmb\bmc]} + \Sigma_{[\bma}{}^{\bmf}{}_{\bmb} R^{\bmd}{}_{|\bme\bm|\bmc]\bmf} =0.\label{2ndBianchiId}
\end{eqnarray}
Furthermore, we recall that the Riemann tensor admits the \emph{irreducible decomposition}
\begin{eqnarray}
&& R^{\bmc}{}_{\bmd\bma\bmb} = C^{\bmc}{}_{\bmd\bma\bmb} + 2 (\delta^{\bmc}{}_{[\bma}L_{\bmb]\bmd} -
   \eta_{\bmd[\bma}L_{\bmb]}{}^{\bmc}), \label{RiemannDecomposition}
\end{eqnarray}
with $C^{\bmc}{}_{\bmd\bma\bmb}$ the components of the \emph{Weyl tensor} and 
\begin{equation}
L_{\bma\bmb} \equiv R_{\bma\bmb} -\frac{1}{6}R \eta_{\bma\bmb}
\label{Definition:Schouten}
\end{equation}
denotes the components of the \emph{Schouten tensor}. The connection
$\mathbf{\nabla}$ is called the \textit{Levi-Civita connection} of $\bmg$ if it
satisfies \eqref{metricComp} and \eqref{torsionFree}. In what follows we
will assume the connection to be Levi-Civita.

\section{The Einstein equations}
\label{section2}
In this work we consider the Einstein equations
\begin{equation}
\label{EFE}
R_{\bma\bmb} - \frac{1}{2}\eta_{\bma\bmb}R = \kappa T_{\bma\bmb} 
\end{equation}
with energy-momentum tensor on the form
\begin{equation}
\label{EnergyMom}
T_{\bma\bmb} = \rho u_{\bma}u_{\bmb}  + \Pi_{\bma\bmb}.
\end{equation}
where $\rho$ is a positive
function of the matter fields. 
We require $\Pi_{\bma\bmb}$ to be a symmetric and purely spatial tensor ---i.e.
\begin{subequations}
\begin{eqnarray}
&&\Pi_{\bma\bmb}u^{\bma} = 0 \label{PiSpatial},\\
&&\Pi_{\bma\bmb} = \Pi_{(\bma\bmb)}. \label{PiSym}
\end{eqnarray}
\end{subequations} 
We do not put any further restrictions on $\Pi_{\bma\bmb}$ other than
it satisfies the \textit{divergence-free condition} of \eqref{EnergyMom}
\begin{equation}
\label{DivEnergyMom}
\nabla^{\bma}T_{\bma\bmb} = 0.
\end{equation}

\begin{remark}\em{
An energy momentum tensor of the form given in \eqref{EnergyMom} is of a very general form and the conditions \eqref{PiSpatial}, \eqref{PiSym} are not stringent restrictions. Thus, the power of the formalism developed herein lies in its generality: given an equation for $\rho$ in terms of the matter fields, one can ignore the matter specific equations of motion and instead solve equations for $\Pi_{\bma\bmb}$. The equations obtained will then be symmetric hyperbolic.} This assumes that one can extract the complete set of equations of motion for the matter fields from \eqref{DivEnergyMom}. 
\end{remark}

\subsubsection*{A projection formalism}
At each point in the spacetime manifold $\mathcal{M}$ the flow lines give rise to a tangent
space which can be split into parts in the direction of $\bmu$
and those orthogonal. This means that without implying a foliation, we
may decompose every tensor defined at each point $p \in \mathcal{M}$
into its orthogonal and timelike part. This may be done by contracting
with $\mathbf{u}$ and the \textit{projector} defined as
\[
h_{\bma}{}^{\bmb} \equiv \eta_{\bma}{}^{\bmb} + u_{\bma}u^{\bmb}, \qquad \bmu = u^{\bma}\mathbf{e}_{\bma}.
\]
Thus, a tensor $T_{\bma\bmb}$ may be split into its time-like, mixed
and space-like parts given, respectively, by
\[
T_{\bm0\bm0}= u^{\bma}u^{\bmb}T_{\bma\bmb}, \qquad T'_{\bm0\bmc}= u^{\bma}h^{\bmb}{}_{\bmc}T_{\bma\bmb}, \qquad T'_{\bmc\bmd}= h^{\bma}{}_{\bmc}h^{\bmb}{}_{\bmd}T_{\bma\bmb},
\]
where $'$ denotes that the free indices left are spatial ---e.g. $T'_{\bma\bm0} u^{\bma} = 0$. Decomposing $\mathbf{\nabla u}$ we
obtain
\begin{equation}
\label{Der4VelDecomp}
\nabla_{\bma} u^{\bmb} = \chi_{\bma}{}^{\bmb} + u_{\bma}a^{\bmb},
\end{equation}
where $\chi_{\bma}{}^{\bmb}$ and $a^{\bmb}$ are the components of the
\textit{Weingarten tensor} and 4-acceleration, respectively, defined
by
\[
\chi_{\bma}{}^{\bmb} \equiv h_{\bma}{}^{\bmc}\nabla_{\bmc} u^{\bmb}, \qquad a^{\bmb} \equiv u^{\bmc}\nabla_{\bmc}u^{\bmb}.
\]
In the literature (e.g. see \cite{Wald1984} p.217) the trace,
trace-free and antisymmetric part of \eqref{Der4VelDecomp} is called,
respectively, the expansion, shear and the twist of the fluid. By
decomposing \eqref{DivEnergyMom} we obtain an equivalent system of
equations in terms of the above quantities
\begin{subequations}
\begin{eqnarray}
&& \nabla^{\bma}\Pi_{\bma\bmb} = -a_{\bmb} \rho + u_{\bmb}\Pi_{\bma\bmc}\chi^{\bma\bmc},\label{DivStress}\\
&&u^{\bma}\nabla_{\bma}\rho = -\rho \chi - \Pi_{\bma \bmb}\chi^{\bma\bmb}.\label{LieRho}
\end{eqnarray}
\end{subequations} 
The decomposition of the \textit{four volume} is
\[
\epsilon_{\bma\bmb\bmc\bmd} = -2\left(u_{[\bma}\epsilon_{\bmb]\bmc\bmd}-\epsilon_{\bma\bmb[\bmc}u_{\bmd]}\right), \qquad \epsilon_{\bmb\bmc\bmd}=\epsilon_{\bma\bmb\bmc\bmd} u^{\bma}.
\]
Given a tensor $T_{abc}$ which is antisymmetric in its two last indices,
we may construct the \textit{electric} and \textit{magnetic} parts with respect to $\mathbf{u}$. In frame indices this is, respectively, defined by
\[
E_{\bmc\bmd} \equiv T_{\bma\bmb\bme} h_{\bmc}{}^{\bma} h_{\bmd}{}^{\bmb} u^{\bme}, \qquad B_{\bmc\bmd} \equiv T^{\ast}{}_{\bma\bmb\bme} h_{\bmc}{}^{\bma} h_{\bmd}{}^{\bmb} u^{\bme},
\]
where the \textit{Hodge dual operator}, denoted by ${}^{\ast}$, is defined
by 
\[
T^{\ast}{}_{\bma\bmb\bme} \equiv -\frac{1}{2}\epsilon^{\bmm\bmn}{}_{\bmb\bme} T_{\bma\bmm\bmn},
\]
and has the property that
\[
T^{\ast \ast}{}_{\bma\bmb\bmc} = -T_{\bma\bmb\bmc}.
\]
Depending on the symmetries and rank of the tensor, the above
definition for electric and magnetic decomposition may vary
slightly. Central for our discussion is that $E_{\bma\bmb}$ and $B_{\bma\bmb}$
are spatial and symmetric.

\section{Gauge considerations}
\label{section3}
\label{Section:GaugeConsiderations}

The gauge to be considered in our hyperbolic reduction procedure for
the Einstein field equations follows the same considerations as in \cite{Friedrich1998}. In
particular, we make the following choices:

\begin{itemize}
\item[{\em i.}] \textbf{\em Orientation of the frame.} We align the
time-leg of the frame with the the flow vector $\mathbf{u}$ tangent to the
worldlines of the particle ---that is, we set
\[
\bmu = \bme_{\bm0}.
\]
\item[\em{ii.}] \textbf{\em Basis in a coordinate system.} Given a
coordinate system $x=(x^\mu)$ we expand the basis vectors as
\begin{equation}
\bme_{\bma} = e_{\bma}{}^\mu \bm\partial_{\mu}.
\label{ExpansionFrame}
\end{equation}
Given an initial hypersurface, $\mathcal{S}_\star$, then the
coordinates $(x^{j})$ defined on $\mathcal{S}_\star$ remain constant along the
flow and, thus, specify the frame. 

\item[\em{iii.}] \textbf{\em Lagrangian condition.} The implementation
of a Lagrangian gauge is equivalent to requiring that
$\mathbf{e}_{\mathbf{0}} = \bm\partial_t$ where $t$ is a suitable
parameter along the world-lines of the material ---e.g. the
\emph{proper time}. In terms of the components of the frame, this
condition is equivalent to requiring that
\begin{equation}
e_{\bm0}{}^{\mu} = \delta_{\bm0}{}^{\mu}.
\label{LagrangianGaugeCondition}
\end{equation}

\item [\em{iv.}] \textbf{\em Fermi Propagation of the frame.} We
require the vector fields $\bme_{\bma}$ to be \textit{Fermi
propagated} along the direction of $\bme_{\bm0}$ ---i.e. 
\[
\nabla_{\bm0} \bme_{\bma} + \bmg\left(\bme_{\bma}, \nabla_{\bm0}\bme_{\bma}\right) \bme_{\bm0} - \bmg\left(\bme_{\bma},  \bme_{\bm0}\right) \nabla_{\bm0}\bme_{\bma} = 0.
\]
This implies the
following conditions on the connection coefficients:
\begin{subequations}
\begin{eqnarray}
&&\Gamma_{\bm0}{}^{\bmi}{}_{\bmj} = 0, \label{FermiProp1}\\
&&\Gamma_{\bm0}{}^{\bm0}{}_{\bm0} = 0, \label{FermiProp2}
\end{eqnarray}
\end{subequations}
for $\bmi,\bmj = 1,\,2,\, 3$.  The second condition is a consequence
of the metric compatibility condition. A frame satisfying the above
equation is a frame where $\bme_{\bm0} = \bmu$ and $\{\bme_{\bmi}\}$ is
orthonormal at every point along the trajectory for which $\bmu$ is
the tangent vector.
\end{itemize}

\section{Zero-quantities}
\label{section4}
In the subsequent discussion it will prove
convenient to introduce, as a book-keeping device, the \emph{zero-quantities}
\begin{subequations}
\begin{eqnarray}
&&\Delta^{\bmd}{}_{\bma\bmb\bmc} \equiv \hat{R}^{\bmd}{}_{\bma\bmb\bmc} - \rho^{\bmd}{}_{\bma\bmb\bmc},\label{DeltaDefinition}\\
&& F_{\bmb\bmc\bmd} \equiv
   \nabla_{\bma}F^{\bma}{}_{\bmb\bmc\bmd}, \label{FriedrichDefinition}\\
&& N_{\bmc\bma\bmb} \equiv Z_{\bmc\bma\bmb} - 2\nabla_{[\bma}\Pi_{\bmb]\bmc},\label{NDef},
\end{eqnarray}
\end{subequations}
where $L_{\bmc\bme}$ denotes the components of the
Schouten tensor as defined by equation \eqref{Definition:Schouten} and $\Pi \equiv \Pi^{\bma}{}_{\bma}$. Moreover, by  $\hat{R}^\bmd{}_{\bma\bmb\bmc}$ it is understood the expression for the Riemann tensor in terms of the connection coefficients $\Gamma_\bma{}^\bmb{}_\bmc$ and its frame derivatives. We have also defined
\begin{subequations}
\begin{eqnarray}
&& \rho^{\bmd}{}_{\bma\bmb\bmc} \equiv \hat{C}^{\bmd}{}_{\bma\bmb\bmc} +  2\eta^{\bmd}{}_{[\bmb}\hat{L}_{\bmc]\bma} -
   2\eta_{\bma[\bmb}\hat{L}_{\bmc]}{}^{\bmd}, \label{rhoDef}\\
&& F^{\bmc}{}_{\bma\bmb\bmd} \equiv \hat{C}^{\bmc}{}_{\bma\bmb\bmd} - 2\eta^{\bmc}{}_{[\bmb}\hat{L}_{\bmd]\bma},\label{FriedrichDef}\\
&& Z_{\bmc\bma\bmb} \equiv 2\nabla_{[\bma}\Pi_{\bmb]\bmc}, \label{ZeldaDef}\\
&&\hat{L}_{\bma\bmb} \equiv T_{\bma\bmb} - \frac{1}{3}\eta_{\bma\bmb}T,\label{ShoutenT},
\end{eqnarray}
\end{subequations}
where $\hat{C}^\bmd{}_{\bma\bmb\bmc}$ is defined as having the same symmetries as the components of the Weyl tensor $C^\bmd{}_{\bma\bmb\bmc}$. 


\begin{remark}
{\em The components $\rho \indices{^{\bmd}_{\bma\bmb\bmc}}$ are
known as the \emph{algebraic curvature} and encode the
decomposition of the Riemann curvature tensor in terms of the Weyl and
Schouten tensors while $F \indices{^{\bmc}_{\bma\bmb\bmc}}$
are the components of the \emph{Friedrich tensor}. The latter provides
a convenient way to encode the second Bianchi identity for the
curvature. }
\end{remark}

\begin{remark}
{\em The tensor $Z_{\bmc\bma\bmb}$, hereafter to be referred to as the \textit{Z-tensor}, is introduced in order for the evolution equations of the electric and magnetic part of the Weyl tensor to be expressed in terms of lower order terms ---i.e. preventing any derivatives of $\Pi_{\bma\bmb}$ to appear in the equations and hence keeping their hyperbolicity.}
\end{remark}

In terms of the objects introduced in the previous paragraphs, the
Einstein field equations \eqref{EFE} can be encoded in the conditions
\begin{subequations}
\begin{eqnarray}
&&\nabla^{\bma} T_{\bma\bmb} = 0, \label{DivergenceFreeCond}\\
&&\Sigma_{\bma}{}^{\bme}{}_{\bmb} = 0, \label{SigmaEquation}\\
&&\Delta\indices{^{\bmd}_{\bma\bmb\bmc}} = 0, \label{DeltaEquation}\\
&& F\indices{_{\bmb\bmc\bmd}} = 0. \label{FriedrichEquation}
\end{eqnarray}
\end{subequations}
More precisely, one has the following result:

\begin{lemma}
\label{Lemma:SolutionFrameSolutionEinstein}
For a given $\rho$, let $(\hat{L}_{\bma\bmb}$, $e^{\mu}{}_{\bma}$, $\Gamma_{\bma}{}^{\bmc}{}_{\bmb}$, $\hat{C}^{\bmd}{}_{\bma\bmb\bmc})$ be a solution to equations \eqref{DivergenceFreeCond}-\eqref{FriedrichEquation} for which the metric compatibility condition  \eqref{metricComp} holds. Then $(\hat{L}_{\bma\bmb}$, $e^{\mu}{}_{\bma}$, $\Gamma_{\bma}{}^{\bmc}{}_{\bmb}$, $\hat{C}^{\bmd}{}_{\bma\bmb\bmc})$ implies the existence of a metric $\bmg$ solution to the Einstein field equations \eqref{EFE} with energy-momentum tensor defined by the components $T_{\bma\bmb}$. Moreover, the fields $\hat{C}^\bmd{}_{\bma\bmb\bmc}$ are, in fact, the components of the Weyl tensor of $\bmg$.
\end{lemma}
\begin{remark}
\em{Note that equations \eqref{DivergenceFreeCond}-\eqref{FriedrichEquation} do not provide a closed system of evolution equations for the unknowns of our system. They are only the necessary equations for giving Lemma \ref{Lemma:SolutionFrameSolutionEinstein}.}
\end{remark}

\begin{proof}
The frame $\{ \mathbf{e}_\bma \}$ obtained from the solution to equation \eqref{SigmaEquation} implies, in turn, by the condition $\langle \bmomega^\bmb, \mathbf{e}_\bma \rangle =\delta_\bma{}^\bmb$ the existence of a coframe $\{\bmomega^{\bmb}\}$ from which one can construct a metric tensor $\bmg$ via the relation 
\[
\bmg = \eta_{\bma\bmb}\bmomega^{\bma} \otimes \bmomega^{\bma}.
\]
Since the coefficients $\Gamma_{\bma}{}^{\bmc}{}_{\bmb}$ satisfy the no-torsion and metric compatibility conditions \eqref{SigmaEquation} and \eqref{metricComp}, then they must coincide with the connection coefficients of the metric $\bmg$ with respect to the frame $\{ \mathbf{e}_\bma \}$. Moreover, by equation \eqref{RiemannExpansion} we have that
\[
\hat{R}^{\bmd}{}_{\bma\bmb\bmc} = R^{\bmd}{}_{\bma\bmb\bmc},
\]
where $R^{\bmd}{}_{\bma\bmb\bmc}$ denotes the frame components of the Riemann curvature tensor. Using the Riemann decomposition as defined by equation \eqref{RiemannDecomposition} together with equation \eqref{DeltaEquation} we obtain
\begin{equation}
C^{\bmd}{}_{\bma\bmb\bmc} +  2\eta^{\bma}{}_{[\bmb}L_{\bmc]\bma} -
   2\eta_{\bma[\bmb}L_{\bmc]}{}^{\bma} = \hat{C}^{\bmd}{}_{\bma\bmb\bmc} +  2\eta^{\bma}{}_{[\bmb}\hat{L}_{\bmc]\bma} -
   2\eta_{\bma[\bmb}\hat{L}_{\bmc]}{}^{\bma}\label{Proof1Eq1}.
\end{equation}
Taking the trace of equation \eqref{Proof1Eq1} with respect to the indices $\bmb$ and $\bmd$ and using the trace-free property of the Weyl tensor and $\hat{C}^\bmd{}_{\bma\bmb\bmc}$  we obtain
\begin{equation}
L_{\bmc\bma} +\frac{1}{2}\eta_{\bmc\bma}L^{\bmd}{}_{\bmd} = \hat{L}_{\bmc\bma} +\frac{1}{2}\eta_{\bmc\bma}\hat{L}^{\bmd}{}_{\bmd}.\label{proof1Eq2}
\end{equation}
Finally, taking the trace of equation \eqref{proof1Eq2} and using equations \eqref{DivEnergyMom} and \eqref{2ndBianchiId}, we get the identity,
\[
L^{\bmd}{}_{\bmd} = \hat{L}^{\bmd}{}_{\bmd}.
\]
The latter shows that $\hat{L}_{\bma\bmb}$ are, in fact, the components of the Schouten tensor of the metric $\bmg$. Using the definition of the Schouten tensor in terms of the Ricci tensor, equation \eqref{Definition:Schouten}, it follows readily that the metric $\bmg$ satisfies the Einstein field equations with an energy-momentum tensor defined by the components $T_{\bma\bmb}$. Returning to equation \eqref{proof1Eq2} we conclude by the uniqueness of the decomposition of the Riemann tensor that the fields $\hat{C}^\bmd{}_{\bma\bmb\bmc}$ are, in fact, the components of the Weyl tensor of $\bmg$.  
\end{proof}

\begin{remark}
{\em In the following to ease the notation, and in a slight abuse of notation we simply write $C^\bmd{}_{\bma\bmb\bmc}$ instead of $\hat{C}^\bmd{}_{\bma\bmb\bmc}$.}
\end{remark}


\section{Evolution equations}


\label{section5}
Given the gauge conditions introduced in Section \ref{Section:GaugeConsiderations}, the next
step in our analysis involves the extraction of a suitable (symmetric
hyperbolic) evolution system from equations
\eqref{DivergenceFreeCond}-\eqref{FriedrichEquation}. We do this in a
number of steps.

\subsection{Equations for the components of the frame}
 The evolution equations for the components of the frame
$e_{\bma}{}^\mu$ are obtained from the \emph{no-torsion condition}
\eqref{SigmaEquation}. In order to do so we exploit the freedom
available in the choice of the frame and require it to be
adapted to the world-lines of the material particles and the gauge conditions
outlined above.

\medskip
Making use of the expansion \eqref{ExpansionFrame} in equation \eqref{SigmaEquation} one readily finds that
\[
e_{\bma}{}^{\mu}\partial_{\mu}e_{\bmb}{}^{\nu} - e_{\bmb}{}^{\mu}\partial_{\mu} e_{\bma}{}^{\nu} =  \left(\Gamma_{\bma}{}^{\mathbf{c}}{}_{\bmb} - \Gamma_{\bmb}{}^{\mathbf{c}}{}_{\bma}\right) e_{\mathbf{c}}{}^{\nu}.
\]
Setting $\bma=0$ in the above expression and making use of the
Lagrangian gauge condition \emph{(iii)} we obtain 
\begin{equation}
\partial_{\mathbf{0}} e_{\bmb}{}^{\nu}  - \left(\Gamma_{\mathbf{0}}{}^{\mathbf{c}}{}_{\bmb} - \Gamma_{\bmb}{}^{\mathbf{c}}{}_{\mathbf{0}}\right) e_{\mathbf{c}}{}^{\nu} =0.
\label{EvolutionEquationCoefficientsFrame}
\end{equation}
This last equation will be read as an evolution equation for the frame
coefficients $e_{\bmb}{}^{\nu}$ with $\bmb=1,\,2,\,3$. As it only
contains derivatives along the flow lines of the matter, it
is, in fact, a transport equation along the world-lines. Observe that
for $\bmb=0$ the equation is satisfied automatically ---recall that as
a consequence of the Lagrangian condition
\eqref{LagrangianGaugeCondition} the coefficients
$e_{\mathbf{0}}{}^\mu$ are already fixed.

\begin{remark}
\label{remarkSigma}
{\em Assuming that the gauge conditions \emph{(i)}, \emph{(ii)} and \emph{(iii)} above hold, equation \eqref{EvolutionEquationCoefficientsFrame} can be succinctly written as
\[
\Sigma_{\mathbf{0}}{}^{\mathbf{c}}{}_{\bmb}=0.
\]
This observation will be of use in the discussion of the propagation of the constraints.}
\end{remark}

\subsection{Evolution equations for the connection coefficients}
The evolution equations for the frame components are given in terms of
the frame connection coefficients. Due to the Fermi propagation and the metric compatibility, equation  \eqref{metricComp}, the independent, non-zero
components of the connection coefficients  are
$\Gamma_{i}{}^{k}{}_{j}$, $\Gamma_{0}{}^{0}{}_{j}$ and
$\Gamma_{i}{}^{0}{}_{j}$. Evolution
equations for $\Gamma_{i}{}^{k}{}_{j}$ may be extracted from the equation for the
algebraic curvature \eqref{DeltaEquation}. More precisely, we consider
the condition
\[
\Delta^{\bmd}{}_{\bmc\bma\bmzero}=0,
\]
which implies
\[
\hat{R}^{\bmd}{}_{\bmc\bma\bmzero} = \rho^{\bmd}{}_{\bmc\bma\bmzero}.
\]
The Riemann tensor can be expressed in terms of the connection
coefficients via equation \eqref{RiemannExpansion}. Furthermore, using
equation \eqref{rhoDef} and the gauge condition \emph{(iv)}, we obtain
\begin{equation}
\begin{split}
\label{EvolutionEquarionCoefficientsConnection}
\partial_{\mathbf{0}} \Gamma_{\bmi}{}^{\bmj}{}_{\bmk} &= - \Gamma_{\bml}{}^{\bmj}{}_{\bmk}\Gamma_{\bmi}{}^{\bml}{}_{\mathbf{0}} - \Gamma_{\mathbf{0}}{}^{\bmj}{}_{\mathbf{0}}\Gamma_{\bmi}{}^{\mathbf{0}}{}_{\bmk} + \Gamma_{\bmi}{}^{\bmj}{}_{\mathbf{0}}\Gamma_{\mathbf{0}}{}^{\mathbf{0}}{}_{\bmk} - C\indices{^{\bmj}_{\bmk\bmi\bmzero}}, 
\end{split}
\end{equation}
where $\bmi,\, \bmj,\, \bmk,\ldots = 1,\,2,\,3$. In the above calculation we have used that  $\Pi_{\bma\bmzero} =  0$ and $\eta^{\bmi}{}_{\bm0} = 0$.
\begin{remark}
\label{Remark:VanishingDelta}
{\em Assuming that the gauge condition \emph{(iv)} holds, equation \eqref{EvolutionEquarionCoefficientsConnection} is equivalent to
\[
\Delta\indices{^{\bmc}_{\bma\bmb\bmzero}}=0.
\]
Observe again, that the resulting equations are, in fact, transport
equations along the world-line of the material particles.}
\end{remark}

The evolution equations for the remaining connection coefficients will be obtained by splitting $\Pi_{\bma\bmb}$ into its trace and trace-free part,
\[
\Pi_{\bma\bmb} = \Pi_{\{\bma\bmb\}} + \frac{1}{3}\Pi \eta_{\bma\bmb},
\]
where $\Pi_{\{\bma\bmb\}}$ denotes the trace-free part of $\Pi_{\bma\bmb}$. Plugging this into \eqref{DivStress} and \eqref{LieRho}, we obtain
\begin{subequations}
\begin{eqnarray}
&& \nabla^{\bma}\Pi_{\{\bma\bmb\}} = -\frac{1}{3}\nabla_{\bmb}\Pi - \rho a_{\bmb} + u_{\bmb}\Pi_{\{\bma\bmc\}}\chi^{\bma\bmc} + \frac{1}{3}\Pi \chi u_{\bmb},\label{DivStressTraceFree}\\
&& u^{\bma}\nabla_{\bma} \rho = -\rho \chi - \Pi_{\{\bma\bmb\}}\chi^{\bma\bmb} - \frac{1}{3}\Pi\chi. \label{LieRhoTraceFree}
\end{eqnarray}
\end{subequations} 
Since $\nabla_{[\bmd}\nabla_{\bmb]}\Pi = 0$, we obtain from equation
\eqref{DivStressTraceFree} that 
\[
J_{\bmd\bmb} = 0,
\]
with
\[
\begin{split}
J_{\bmd\bmb} &\equiv -2\rho\nabla_{[\bmb}a_{\bmd} + 2a_{[\bmd}\nabla_{\bmb]}\rho + 2\Pi_{\{\bma\bmc\}}\chi^{\bma\bmc}\nabla_{[\bmd}u_{\bmb]} + 2u_{[\bmb}\nabla_{\bmd]}\left(\Pi_{\{\bma\bmc\}}\chi^{\bma\bmc}\right)\\
&+ \frac{2}{3}\Pi\chi\nabla_{[\bmd}u_{\bmb]} + \frac{2}{3}u_{[\bmb}\nabla_{\bmd]}\left(\Pi\chi\right) - 2\nabla_{[\bmd}\nabla^{\bma}\Pi_{\{\bmb]\bma\}}.
\end{split}
\]
The last term may be written,
\begin{equation}
\label{LastTerminJ}
\begin{split}
\nabla_{[\bmd}\nabla^{\bma}\Pi_{\{\bmb]\bma\}} &= -R^{\bmm}{}_{\bmb\bmd}{}^{\bma}\Pi_{\{\bmm\bma\}} + R^{\bmm}{}_{\bmd\bmb}{}^{\bma}\Pi_{\{\bmm\bma\}} +2R^{\bmm}{}_{[\bmd}\Pi_{\{\bma]\bmm\}} + \nabla^{\bma}Z_{\bma\bmd\bmb},\\
&= 2R^{\bmm}{}_{[\bmd}\Pi_{\{\bma]\bmm\}} + \nabla^{\bma}Z_{\bma\bmd\bmb},
\end{split}
\end{equation}
where we have used the symmetry of the Riemann tensor in the last step. A straight forward calculation shows that,
\[
Z_{\bma\bmd\bmb} = \nabla_{\bmm}C^{\bmm}{}_{\bma\bmd\bmb} + P_{\bma\bmd\bmb},
\]
where,
\[
P_{\bma\bmd\bmb} \equiv \frac{2}{3}\nabla_{[\bmd}(\eta_{\bmb]\bma}T) -2\nabla_{[\bmd}(\rho u_{\bmb]}u_{\bma}).
\]
Using that,
\[
\nabla_{\bma}\nabla_{\bmb}C^{\bma\bmb}{}_{\bmc\bmd}=0, \qquad \nabla_{[\bma}\nabla_{\bmb]}T = 0,
\]
we have,
\begin{equation}
\label{DivergenceZelda}
\nabla^{\bma}Z_{\bma\bmd\bmb} =-2  \nabla^{\bma}\nabla_{[\bmd}(\rho u_{\bmb]}u_{\bma}).
\end{equation}
Substituting this back into \eqref{LastTerminJ}, we may now write the $\{0,\bmi\}$ components of $J_{\bmd\bmb}$ as
\begin{equation}
\label{JoiCompExp}
J_{\bm0\bmi}=-2\rho \nabla_{\bm0}a_{\bmi} - R_{\bm0}{}^{\bmj} \Pi_{\bmi\bmj} - 2 a_{\bmi} \rho \chi - 2 a_{\bmi} \Pi_{\bmi\bmj} \chi^{\bmi\bmj}  + a^{\bmj} \rho \chi_{\bmi\bmj} -  \rho \nabla_{\bmj}\chi_{\bmi}{}^{\bmj} +
\rho \nabla_{\bmi}\chi.
\end{equation}
In the above expression we have used the Lagrangian gauge condition to set $u_{\bmi} = 0$. 
Finally, using the definition for $a_{\bmi}$, we readily obtain 
\begin{equation}
\begin{split}
3\partial_{\bm0} \Gamma_{\bm0}{}^{\bm0}{}_{\bmi} -
\eta^{\bmj\bmk}\partial_{\bmi}\Gamma_{\bmj}{}^{\bm0}{}_{\bmk} &= - 2a_{\bmi}\chi + a^{\bmj}\chi_{\bmi\bmj} - \partial^{\bmj}\Gamma_{\bmj}{}^{\bm0}{}_{\bmi} +\Gamma_{\bmj}{}^{\bmk}{}_{\bmi}\chi_{\bmk}{}^{\bmj} -\Gamma_{\bmj}{}^{\bmj}{}_{\bmk}\chi_{\bmi}{}^{\bmk}\\
&- \frac{1}{\rho}\left(R_{\bm0}{}^{\bmj} \Pi_{\bmi\bmj}  + 2 a_{\bmi} \Pi_{\bmi\bmj} \chi^{\bmi\bmj}\right) - \Gamma_{\bmj}{}^{\bm0}{}_{\bm0}\chi^{\bmj}{}_{\bmi}- \Gamma_{\bm0}{}^{\bm0}{}_{\bmi}\chi\\
&+ \Gamma_{\bm0}{}^{\bmj}{}_{\bmi}\Gamma_{\bm0}{}^{\bm0}{}_{\bmj} + \eta^{\bmj\bmk}\Gamma_{\bmi}{}^{\bml}{}_{\bmj}\Gamma_{\bml}{}^{\bm0}{}_{\bmk} + \eta^{\bmj\bmk}\Gamma_{\bmi}{}^{\bml}{}_{\bmk}\Gamma_{\bmj}{}^{\bm0}{}_{\bml}. \label{evolConnection2}
\end{split}
\end{equation}
To obtain the evolution equation for the remaining connection coefficient, we consider the $\{\bmi,\bmj\}$ components of $J_{\bma\bmb}$:
\begin{equation}
\label{JijComp}
\begin{split}
\partial_ {\bm0}\Gamma_{\bmi}{}^{\bm0}{}_{\bmj} &= \frac{1}{\rho}2R_{\bmj}{}^{\bmk} \Pi_{\bmi\bmk} - \frac{1}{\rho}R_{\bmi}{}^{\bmk} \Pi_{\bmj\bmk} - \chi_{\bmk\bmj} \chi_{\bmi}{}^{\bmk} + \chi_{\bmk\bmi}\chi_{\bmj}{}^{\bmk} + \partial_ {0}\Gamma_{\bmj}{}^{\bm0}{}_{\bmi} + \partial_{\bmi}\Gamma_{\bm0}{}^{\bm0}{}_{\bmj}\\
&+ \frac{1}{\rho}2 a_{\bmj} \nabla_{\bmi}\rho - \partial_{\bmj}\Gamma_{\bm0}{}^{\bm0}{}_{\bmi} - \frac{1}{\rho}2 a_{\bmi} \nabla_{\bmj}\rho - \Gamma_{\bmi}{}^{\bm0}{}_{\bm0}a_{\bmj} + \Gamma_{\bmj}{}^{\bm0}{}_{\bm0}a_{\bmi} + \Gamma_{\bm0}{}^{\bmk}{}_{\bmi}\Gamma_{\bmk}{}^{\bm0}{}_{\bmj}\\
& + \Gamma_{\bm0}{}^{\bmk}{}_{\bmj}\Gamma_{\bmi}{}^{\bm0}{}_{\bmk} - \Gamma_{\bm0}{}^{\bmk}{}_{\bmj}\Gamma_{\bmk}{}^{\bm0}{}_{\bmi} - \Gamma_{\bm0}{}^{\bmk}{}_{\bmi}\Gamma_{\bmj}{}^{\bm0}{}_{\bmk} + \Gamma_{\bm0}{}^{\bmk}{}_{\bmi}\Gamma_{\bm0}{}^{\bm0}{}_{\bmk} - \Gamma_{\bm0}{}^{\bmk}{}_{\bmj}\Gamma_{\bm0}{}^{\bm0}{}_{\bmk}.
\end{split}
\end{equation}
From equation \eqref{RiemannExpansion} we have,
\[
\partial_{\bmj} \Gamma_{\bm0}{}^{\bm0}{}_\bmi = \partial_{\bm0}\Gamma_{\bmj}{}^{\bm0}{}_{\bmi} -R^{\bm0}{}_{\bmi\bm0\bmj} +
   \Gamma_\bmf{}^{\bm0}{}_\bmi(\Gamma_\bmj{}^\bmf{}_{\bm0} -
   \Gamma_{\bm0}{}^\bmf{}_\bmj) +
   \Gamma_\bmj{}^\bmf{}_\bmi\Gamma_{\bm0}{}^{\bm0}{}_\bmf -
   \Gamma_{\bm0}{}^\bmf{}_\bmi\Gamma_\bmj{}^{\bm0}{}_\bmf
   -\Sigma_{\bm0}{}^\bmf{}_\bmj \Gamma_\bmf{}^{\bm0}{}_\bmi.
\]
Substituting this back into \eqref{JijComp} gives the final evolution equation:
\begin{equation}
\label{evolConnection3}
\begin{split}
\partial_ {\bm0}\Gamma_{\bmi}{}^{\bm0}{}_{\bmj}- \partial_{\bmi}\Gamma_{\bm0}{}^{\bm0}{}_{\bmj} &= \frac{2}{\rho}R_{[\bmj}{}^{\bmk} \Pi_{\bmi]\bmk} + 2 \chi_{\bmk[\bmi}\chi_{\bmj]}{}^{\bmk} + \frac{4}{\rho}a_{[\bmj} \nabla_{\bmi]}\rho - R^{\bm0}{}_{\bmi\bm0\bmj}\\
&-\Gamma_\bmk{}^{\bm0}{}_{\bmi}\Gamma_{\bmj}{}^\bmk{}_{\bm0} -\Gamma_\bmj{}^\bmk{}_\bmi\Gamma_{\bm0}{}^{\bm0}{}_{\bmk} - \Gamma_{\bmi}{}^{\bm0}{}_{\bm0}a_{\bmj} + \Gamma_{\bmj}{}^{\bm0}{}_{\bm0}a_{\bmi}\\
&+ \Gamma_{\bm0}{}^{\bmk}{}_{\bmi}\Gamma_{\bmk}{}^{\bm0}{}_{\bmj}+ \Gamma_{\bm0}{}^{\bmk}{}_{\bmj}\Gamma_{\bmi}{}^{\bm0}{}_{\bmk} + \Gamma_{\bm0}{}^{\bmk}{}_{\bmi}\Gamma_{\bm0}{}^{\bm0}{}_{\bmk} - \Gamma_{\bm0}{}^{\bmk}{}_{\bmj}\Gamma_{\bm0}{}^{\bm0}{}_{\bmk}.
\end{split}
\end{equation}
Equations \eqref{evolConnection2} and \eqref{evolConnection3} are on a form which is known to be symmetric hyperbolic --- we refer again to \cite{Alho2010} for details.
\begin{remark}
\em{The presence of a spatial derivative of $\rho$ in equations \eqref{evolConnection2} and \eqref{evolConnection3} means that an equation for $\rho$ is necessary to ensure the hyperbolicity of the equations.}
\end{remark}

\subsection{Evolution equations for the decomposed $Z$-tensor}
It is well known that in vacuum the Bianchi equation leads to a symmetric hyperbolic equation for the independent components of the Weyl tensor. By contrast,
an inspection of the definition of the Friedrich tensor $F_{\bma\bmb\bmc\bmd}$, equation \eqref{FriedrichDefinition}, reveals that the condition $F_{\bma\bmb\bmc}=0$ involves both derivatives of $C_{\bma\bmb\bmc\bmd}$ and the matter variables. This potentially destroys the symmetric hyperbolicity of the equation for the components of the Weyl tensor. In the following we will show that it is possible to deal with this difficulty by providing two auxiliary fields ---the $Z$-tensor and $\sigma$-tensor as defined by equations \eqref{ZeldaDef} and \eqref{SigmaDef}, respectively.


\medskip
We first define some important quantities and identities used in the following discussion. 
The $Z$-tensor has the symmetries
\[
Z_{[\bma\bmb\bmc]} = 0 	,\qquad Z_{\bma\bmb\bmc} = Z_{\bma[\bmb\bmc]} .
\]
The symmetry of the $Z$-tensor thus allows for a decomposition in
terms of its electric and magnetic parts defined respectively as
\[
\Psi_{\bma\bmc} \equiv Z_{\bme\bmb\bmd}u^{\bmd}
h_{\bma}{}^{\bme}h_{\bmc}{}^{\bmb}, 
 \qquad \Phi_{\bma\bmc} \equiv Z^{\ast}_{\bme\bmb\bmd}u^{\bmd}
h_{\bma}{}^{\bme}h_{\bmc}{}^{\bmb},
\]
where, $Z^{\ast}_{\bme\bmb\bmd}$, is the dual $Z$-tensor defined in the customary way.
The electric and magnetic part of the $Z$-tensor are symmetric tensors defined on the orthogonal space of $\bmu$ ---i.e. one has that 
\[
\Psi_{\bma\bmc} = \Psi_{(\bma\bmc)},\qquad
\Psi_{\bma\bmc} u^{\bma}=0, \qquad
\Phi_{\bma\bmc} = \Phi_{(\bma\bmc)},\qquad
\Phi_{\bma\bmc} u^{\bma}=0.
\]
 As such, the $Z$-tensor and its dual may be expressed in terms of the
 spatial fields
 \begin{subequations}
\begin{eqnarray}
 &&Z_{\bmc\bma\bmb} = \Psi_{\bmc\bmb} u_{\bma} -  \Psi_{\bmc\bma} u_{\bmb} -  \epsilon_{\bma\bmb}{}^{\bme}
\Phi_{\bmc\bme} + u_{\bmc} \Pi_{\bmd\bmb} \chi_{\bma}{}^{\bmd} -  u_{\bmc} \Pi_{\bmd\bma}
\chi_{\bmb}{}^{\bmd}, \label{ZeldaElectricDecomp}\\
&&Z^{\ast}{}_{\bma\bmm\bmn} = \tfrac{1}{2} \Psi_{\bma\bmc} \epsilon_{\bmm\bmn}{}^{\bmc} + u_{[\bmm} \Phi_{\bmn]\bma}
+ \tfrac{1}{2} \epsilon_{\bmm\bmn\bmd} \Pi_{\bmc}{}^{\bmd} \chi_{\bma}{}^{\bmc} -  \tfrac{1}{2}\epsilon_{\bmm\bmn\bmc} \Pi_{\bma\bmd} \chi^{\bmc\bmd}. \label{ZeldaMagneticDecomp}
\end{eqnarray}
\end{subequations}
By plugging the definition for the $Z$-tensor into the definitions of $\Psi_{\bma\bmb}$ and
$\Phi_{\bma\bmb}$, respectively, we obtain an evolution equation for the
matter tensor $\Pi_{\bma\bmb}$ in terms of $\Psi_{\bma\bmb}$ together with a constraint equation. Namely, one has that 
\begin{subequations}
\begin{eqnarray}
 && u^{\bma} \nabla_{\bma}\Pi_{\bmf\bmm} = a^{\bma} u_{\bmm} \Pi_{\bmf\bma} + a^{\bma} u_{\bmf} \Pi_{\bmm\bma} -  
\Pi_{\bmm\bma} \chi_{\bmf}{}^{\bma} -\Psi_{\bmf\bmm},\label{EvolMatter1} \\
&&\epsilon_{\bmf}{}^{\bmb}{}_{\bma} \mathcal{D}_{\bmb}\Pi_{\bmm}{}^{\bma} = \epsilon_{\bmf\bmb\bma} u_{\bmm} \Pi_{\bmc}{}^{\bma} \chi^{\bmb\bmc}  
 +\Phi_{\bmf\bmm}\label{EvolMatter2}.
\end{eqnarray}
\end{subequations}
where $\mathcal{D}_{\bmb}$ denotes the \textit{Sen connection} defined as,
\[
\mathcal{D}_{\bmb} \Pi_{\bmc\bmd} \equiv h_{\bmb}{}^{\bma}\nabla_{\bma}\Pi_{\bmc\bmd}.
\]
It is worth noting that due to the $1 +3$ split of space time, we do not have a spatial metric on the 3-dimensional hypersurfaces. Hence, we cannot define a spatial derivative --- i.e. a spatial metric satisfying the metric compatibility condition does not exist on the three surfaces. 
Equation \eqref{EvolMatter2} is regarded as a constraint equation.
\begin{remark}
\em Note that equation \eqref{EvolMatter2} will always hold as long as the definition of the $Z$-tensor (i.e. equation \eqref{ZeldaDef}) propagates. This will be showed in Section \ref{section6}.
\end{remark}
In order to close the system and to ensure hyperbolicity a set of evolution equations for the fields $\Psi_{\bma\bmb}$ and $\Phi_{\bma\bmb}$ are needed. In the rest of this section we shall develop these equations and show they form a  first order symmetric hyperbolic system.

\smallskip
The evolution equations for $\Psi_{\bma\bmb}$ is obtained by taking the divergence of the $Z$-tensor ---i.e we have the equation
\[
\nabla^{\bmb}Z_{\bmc\bma\bmb} = 2\nabla^{\bmb}\nabla_{[\bma}\Pi_{\bmb]\bmc}.
\]
Expanding the above equation and using the decomposition of the $Z$-tensor ---i.e. equation \eqref{ZeldaElectricDecomp}--- we obtain after a number of steps the equation

\begin{equation}
\label{PsiEvolutionEq}
u^{\bmb}\nabla_{\bmb}\Psi_{\bma\bmc} - \epsilon_{\bma}{}^{\bmb\bmd}\mathcal{D}_{\bmd}\Phi_{\bmc\bmb} = \mathcal{W}_{\bma\bmc},
\end{equation}
where $\mathcal{W}_{\bma\bmb}$ denotes the lower order terms and is explicitly given by 
\[
\begin{split}
\mathcal{W}_{\bma\bmb} &= -2 a^{\bmb} \Psi_{\bmc\bmb} u_{\bma} -  a^{\bmb} \Psi_{\bma\bmb} u_{\bmc} -  a_{\bmb} a^{\bmb} u_{\bma} u_{\bmc} \rho + a^{\bmb} \epsilon_{\bma\bmb\bmd} \Phi_{\bmc}{}^{\bmd} + a^{\bmb} a^{\bmd} u_{\bma} u_{\bmc} \Omega_{\bmb\bmd} + R_{\bma\bmb\bmc\bmd} \Pi^{\bmb\bmd}\\
&+ R_{\bma}{}^{\bmb} \Pi_{\bmc\bmb} + u_{\bma} u^{\bmb}R_{\bmb}{}^{\bmd} \Pi_{\bmc\bmd} + u^{\bmb} u_{\bmc} R_{\bma\bmd\bmb\bme} \Pi^{\bmd\bme} + 2 u_{\bma}u^{\bmb} R_{\bmc\bmd\bmb\bme} \Pi^{\bmd\bme} + 2 u_{\bma} u^{\bmb} u_{\bmc} u^{\bmd} R_{\bmb\bme\bmd\bmf}
\Pi^{\bme\bmf}\\
&-  a^{\bmb} u_{\bmc} \rho \chi_{\bma\bmb} + a^{\bmb} u_{\bma} \Pi_{\bmc\bmd}\chi_{\bmb}{}^{\bmd} + \Psi_{\bmc\bmb} \chi^{\bmb}{}_{\bma} -  \Psi_{\bma\bmc} \chi^{\bmb}{}_{\bmb} -  a_{\bmc} u_{\bma} \rho \chi^{\bmb}{}_{\bmb} -  \Pi_{\bma\bmd} \chi_{\bmb}{}^{\bmd} \
\chi^{\bmb}{}_{\bmc}\\
&-  \Psi_{\bmb\bmd} u_{\bma} u_{\bmc} \chi^{\bmb\bmd} + \epsilon_{\bmb\bmd\bme} u_{\bma} \Phi_{\bmc}{}^{\bme} \chi^{\bmb\bmd} + \epsilon_{\bmc\bmb\bme} u_{\bma} \Phi_{\bmd}{}^{\bme} \chi^{\bmb\bmd} + \epsilon_{\bma\bmb\bme} u_{\bmc} \Phi_{\bmd}{}^{\bme} \chi^{\bmb\bmd}\\
&-  a_{\bmc} u_{\bma} \Pi_{\bmb\bmd} \chi^{\bmb\bmd} -  \Pi_{\bmb\bmd} \chi_{\bma\bmc} \chi^{\bmb\bmd} - 3u_{\bma} u_{\bmc} \Pi_{\bmd\bme} \chi_{\bmb}{}^{\bme} \chi^{\bmb\bmd} -  a^{\bmb} u_{\bma} \Pi_{\bmb\bmd} \chi_{\bmc}{}^{\bmd} + a^{\bmb} u_{\bma} \Pi_{\bmb\bmd} \chi^{\bmd}{}_{\bmc}\\
&+ \Pi_{\bmb\bmd} \chi_{\bma}{}^{\bmb} \chi^{\bmd}{}_{\bmc} + a^{\bmb} u_{\bma} \Pi_{\bmc\bmb} \chi^{\bmd}{}_{\bmd} + 2 u_{\bma} u_{\bmc} \Pi_{\bmb\bme} \chi^{\bmb\bmd} \chi^{\bme}{}_{\bmd} + \rho \nabla_{\bma}a_{\bmc} + a_{\bmc} \nabla_{\bma}\rho + u_{\bma} u^{\bmb} \rho \nabla_{\bmb}a_{\bmc}\\
&+ u_{\bma} u^{\bmb} \Pi_{\bmc\bmd} \nabla_{\bmb}a^{\bmd} + u_{\bma} u^{\bmb} u_{\bmc} \Pi_{\bmd\bme} \nabla_{\bmb}\chi^{\bmd\bme} + \sigma_{\bma\bmc} + u_{\bma} \Psi_{\bmb\bmd} u_{\bmc} \chi^{\bmb\bmd} -  u_{\bma}\epsilon_{\bmc\bmd\bma} \Phi_{\bmb}{}^{\bma} \chi^{\bmb\bmd} + u_{\bma} \Pi_{\bmb\bmd} \nabla_{\bmc}\chi^{\bmb\bmd}\\
&+ u^{\bmb} u_{\bmc} \sigma_{\bma\bmb} + u_{\bma} u^{\bmb} \sigma_{\bmc\bmb} + u_{\bma} u_{\bmc} \Pi_{\bmb\bmd} \nabla^{\bmd}a^{\bmb} -  u_{\bma} \Pi_{\bmb\bmd} \nabla^{\bmd}\chi_{\bmc}{}^{\bmb} + 2 u_{\bma} u^{\bmb} u_{\bmc} u^{\bmd} \sigma_{\bmb\bmd}.
\end{split}
\]
In the above, we have defined
\[
\sigma_{\bma\bmb} \equiv \nabla^{\bmc}\nabla_{\bmc}\Pi_{\bma\bmb}. \label{SigmaDef}
\]
Note that
the derivatives of $\chi_{\bma\bmb}$ and $a_{\bma}$ may be expressed
in terms of the connection coefficients and thus dealt with by
\eqref{EvolutionEquarionCoefficientsConnection} and the definition for
the Riemann tensor ---i.e. we have
\[
\chi_{\bma\bmb} = h_{\bma}{}^{\bmm} h_{\bmb}{}^{\bmn} \Gamma_{\bmm}{}^{\bm0}{}_{\bmn}, \qquad a_{\bma} = h_{\bma\bmc}\Gamma_{\bm0}{}^{\bmc}{}_{\bm0}.
\]
In obtaining equation \eqref{PsiEvolutionEq} we have used standard tensor manipulations involving the commutation of derivatives using the Riemann tensor and frequently making use of the spatial property of $\Pi_{\bma\bmb}$ to get rid of derivatives. In particular, we have used the  identities
\[
2\chi^{\bma\bmb}\nabla_{\bma} \Pi_{\bmd\bmb} = -\chi^{\bma\bmb}\nabla_{\bmd}\Pi_{\bma\bma} + L.O.T, \qquad \nabla_{\bma}\Pi_{\bmb\bmc} = Z_{\bmc\bma\bmb} + \nabla_{\bmb}\Pi_{\bma\bmc},
\]
where \textit{L.O.T} is a shortening for ``lower order terms". The first is obtained from considering $\nabla^{\bmb}Z'_{\bmc\bm0\bmb}$ and the second is a trivial result of the definition for the $Z$-tensor.
\begin{remark}
\em To preserve the hyperbolicity of equation \eqref{PsiEvolutionEq} it is understood that the tensor $\sigma_{\bma\bmb}$
is given in terms of lower order terms. It can readily be shown that $\sigma_{\bm0\bm0} = L.O.T$ from the temporal part of equation \eqref{PsiEvolutionEq}; with the symmetry of $\sigma_{\bma\bmb}$, it is thus 9 components which need to be specified:
\[
\sigma_{\bma\bmb} = \mathcal{F}\left(L.O.T\right).
\]
\end{remark}

\smallskip
To obtain the evolution equation for the field $\Phi_{\bma\bmb}$, we proceed in a similar way as above by considering the dual equation,
\[
\nabla^{\bmb}Z^{\ast}{}_{\bmc\bma\bmb} = -\epsilon_{\bma\bmb}{}^{\bmd\bme}\nabla^{\bmb}\nabla_{[\bmd}\Pi_{\bme]\bmc}.
\]
By applying the decomposition \eqref{ZeldaMagneticDecomp} and expanding we obtain after a few manipulations the evolution equation on the desired form,
\begin{equation}
\label{EvolEqPhi}
u^{\bmb} \nabla_{\bmb}\Phi_{\bma\bmc}  + \epsilon_{\bma\bmb\bmm} \mathcal{D}^{\bmm}\Psi_{\bmc}{}^{\bmb} = \mathcal{U}_{\bma\bmc},
\end{equation}
with $\mathcal{U}_{\bma\bmc}$ denoting the lower order terms ---explicitly given by
\[
\begin{split}
\mathcal{U}_{\bma\bmc} &=- a^{\bmb} \Psi_{\bmc}{}^{\bmm} \epsilon_{\bma\bmb\bmm} -  a^{\bmb} u_{\bma} \Phi_{\bmc\bmb} + 2\epsilon_{\bmb}{}^{\bmn\bmd} u_{\bma} R_{\bmc\bmm\bmn\bmd} \Pi^{\bmb\bmm} - 2 \epsilon^{\bmm\bmn\bmd}u_{\bma} R_{\bmb\bmm\bmn\bmd} \Pi_{\bmc}{}^{\bmb}\\
&-  \epsilon_{\bma}{}^{\bmn\bmd} u^{\bmb} R_{\bmb\bmm\bmn\bmd} \Pi_{\bmc}{}^{\bmm} -  \epsilon_{\bma}{}^{\bmn\bmd} u^{\bmb} R_{\bmb\bmn\bmm\bmd}\Pi_{\bmc}{}^{\bmm} + \epsilon_{\bma\bmm}{}^{\bmd} u^{\bmb} R_{\bmc\bmn\bmb\bmd} \Pi^{\bmm\bmn} + \Phi_{\bmc\bmb} \chi^{\bmb}{}_{\bma}\\
&-  \Phi_{\bma\bmc} \chi^{\bmb}{}_{\bmb} - 2
\Psi_{\bmc}{}^{\bmn} \epsilon_{\bmb\bmm\bmn} u_{\bma} \chi^{\bmb\bmm} -  \Psi_{\bmm}{}^{\bmn}\epsilon_{\bma\bmb\bmn} u_{\bmc} \chi^{\bmb\bmm}+ 2 u_{\bma} u_{\bmc} \Phi_{\bmb\bmm} \chi^{\bmb\bmm} -
a_{\bmc} \epsilon_{\bma\bmb\bmn} \Pi_{\bmm}{}^{\bmn} \chi^{\bmb\bmm}\\
&-  \epsilon_{\bma\bmb\bmd} u_{\bmc} \Pi_{\bmn}{}^{\bmd} \chi^{\bmb\bmm} \chi_{\bmm}{}^{\bmn} + a^{\bmb}\epsilon_{\bmm\bmn\bmd} u_{\bma} u_{\bmc} \Pi_{\bmb}{}^{\bmd} \chi^{\bmm\bmn} - \epsilon_{\bmb\bmm\bmd} u_{\bmc} \Pi_{\bmn}{}^{\bmd} \chi^{\bmb}{}_{\bma} \chi^{\bmm\bmn}\\
&+\epsilon_{\bma\bmm\bmd} u_{\bmc} \Pi_{\bmn}{}^{\bmd} \chi^{\bmb}{}_{\bmb} \chi^{\bmm\bmn} - 2\epsilon_{\bmb\bmm\bmd} u_{\bma} \Pi_{\bmn}{}^{\bmd} \chi^{\bmb}{}_{\bmc} \chi^{\bmm\bmn} - \epsilon_{\bmm\bmn\bmd} u_{\bma} \Pi_{\bmb}{}^{\bmd} \chi_{\bmc}{}^{\bmb} \chi^{\bmm\bmn}\\ &+\epsilon_{\bmb\bmn\bmd} u_{\bma} \Pi_{\bmc\bmm} \chi^{\bmb\bmm} \chi^{\bmn\bmd} + \epsilon_{\bma\bmm\bmd} u^{\bmb} u_{\bmc} \Pi_{\bmn}{}^{\bmd} \nabla_{\bmb}\chi^{\bmm\bmn} + 2 \epsilon_{\bmb\bmn\bmd} u_{\bma} u_{\bmc}\Pi_{\bmm}{}^{\bmd} \nabla^{\bmn}\chi^{\bmb\bmm}.
\end{split}
\]
Equations \eqref{PsiEvolutionEq} and \eqref{EvolEqPhi} are on a form known to be symmetric hyperbolic --- see \cite{Alho2010} for a more detailed discussion.
We note that we have made ample use of the suite {\tt
xAct}\footnote{See http://www.xact.es for more information.} to obtain $\mathcal{W}$ and $\mathcal{U}$.
\begin{remark}
\em In the expressions for $\mathcal{W}_{\bma\bmc}$ and $\mathcal{U}_{\bma\bmc}$ it is understood that wherever $R^{\bmd}{}_{\bma\bmb\bmc}$ appears, it is to be evaluated using the decomposition in terms of the $C^{\bmd}{}_{\bma\bmb\bmc}$ and $\hat{L}_{\bma\bmb}$. 
\end{remark}
\subsection{Evolution equations for the decomposed Weyl tensor}
The construction of suitable evolution equations for the components of
the Weyl tensor follows a similar approach as in the previous discussion. Again, the strategy is to
decompose the Weyl tensor into parts orthogonal to the 4-velocity
---i.e. one needs to understand the form the Weyl tensor takes on the
orthogonal space to the 4-velocity. 

\smallskip
Due to the symmetries of the Weyl
tensor, the essential components are encoded in what are called the
\textit{electric} and \textit{magnetic parts of the Weyl tensor}
defined, respectively, as
\[
E_{\bma\bmc} \equiv C_{\bme\bmb\bmf\bmd}u^{\bmb}u^{\bmd}
h_{\bma}{}^{\bme}h_{\bmc}{}^{\bmf}, 
 \qquad B_{\bma\bmc} \equiv C^{\ast}_{\bme\bmb\bmf\bmd}u^{\bmb}u^{\bmd} h_{\bma}{}^{\bme}h_{\bmc}{}^{\bmf},
\]
where $C^{\ast}_{\bma\bmb\bmc\bmd}$ denotes components the
\textit{Hodge dual} of the Weyl tensor. In terms of these
spatial tensors, the frame components of the Weyl tensor and its dual
admit the decomposition
\begin{eqnarray*}
&& C_{\bma\bmb\bmc\bmd} =
   -2\left(l_{\bmb[\bmc}E_{\bmd]\bma}-l_{\bma[\bmc}E_{\bmd]\bmb}\right)-2\left(u_{[\bmc}B_{\bmd]\bmp}\epsilon^{\bmp}{}_{\bma\bmb}
   + u_{[\bma}B_{\bmb]\bmp}\epsilon^{\bmp}{}_{\bmc\bmd}\right), \\
&& C^{\ast}_{\bma\bmb\bmc\bmd} = 2u_{[\bma}E_{\bmb]\bmp}\epsilon^{\bmp}{}_{\bmc\bmd} -4E^{\bmp}{}_{[\bma}\epsilon_{\bmb]\bmp[\mathbf{c}}u_{\bmd]}-4u_{[\bma}B_{\bmb][\mathbf{c}}u_{\bmd]} - B_{\bmp\bmq}\epsilon^{\bmp}{}_{\bma\bmb}\epsilon^{\bmq}_{\bmc\bmd},
\end{eqnarray*}
---see e.g. \cite{Kroon2016} for details. For convenience we have written
\[
l_{\bma\bmb} \equiv h_{\bma\bmb} - u_{\bma}u_{\bmb}.
\]
In order to obtain evolution equations for  $E_{\bma\bmb}$ and
$B_{\bma\bmb}$, we make use of the following decomposition of the Bianchi identity
\eqref{FriedrichEquation} and its dual: 
\begin{subequations}
\begin{eqnarray}
&& F_{\bmb\bmc\bmd} = u_{\bmb}\left(F'_{\mathbf{0}\bmc\bmzero}u_{\bmd} -
   F'_{\mathbf{0}\bmd\bmzero}u_{\bmc}\right)+
   2F'_{\bmb\bmzero[\bmc}u_{\bmd]} - u_{\bmb}F'_{\bmzero\bmc\bmd} +
   F'_{\bmb\bmc\bmd}, \label{ReducedBianchi} \\
&& F^{\ast}_{\bmb\bmc\bmd} =
   u_{\bmb}\left(F'^{\ast}_{\mathbf{0}\bmc\bmzero}u_{\bmd} -
   F'^{\ast}_{\mathbf{0}\bmd\bmzero}u_{\bmc}\right)+
   2F'^{\ast}_{\bmb\bmzero[\bmc}u_{\bmd]} -
   u_{\bmb}F'^{\ast}_{\bmzero\bmc\bmd} + F'^{\ast}_{\bmb\bmc\bmd}. \label{ReducedBianchiDual}
\end{eqnarray}
\end{subequations}
The term $F_{\bma\bmzero\bmb} = - F_{\bma\bmb\bmzero}$ in 
equation \eqref{ReducedBianchi} gives the evolution equations for
$E_{\bma\bmb}$. More precisely, after a long computation one finds that 
\begin{equation}
\label{EvolEqElWeyl}
\begin{split}
u^{\bmc} \nabla_{\bmc}E_{\bma\bmb} -\epsilon _{\bma\bme\bmf} D^{\bmf}B_{\bmb}{}^{\bme}  &=  \tfrac{1}{2} \kappa  \Psi _{\bma\bmb} - a^{\bmc} E_{\bmb\bmc} u_{\bma} - a^{\bmc} E_{\bma\bmc} u_{\bmb}\\
&- \tfrac{1}{6} \kappa  \Psi ^{\bmc}{}_{\bmc} h_{\bma\bmb} + a^{\bmc} \epsilon_{\bmc\bme\bmf} B_{\bma}{}^{\bme} h_{\bmb}{}^{\bmf} - \kappa  a^{\bmc} u_{\bmb} \Pi_{\bma\bmc}\\
&- \kappa  a^{\bmc} u_{\bma} \Pi _{\bmb\bmc} + \tfrac{1}{2} \kappa \Pi _{\bmb\bmc} \chi _{\bma}{}^{\bmc} - \tfrac{1}{2} \kappa  \rho  \chi _{\bmb\bma} - E_{\bma\bmc} \chi _{\bmb}{}^{\bmc}\\
&- \tfrac{1}{2} \kappa  \Pi _{\bma\bmc} \chi_{\bmb}{}^{\bmc} + 2 E_{\bmb\bmc} \chi ^{\bmc}{}_{\bma} - 2 E_{\bma\bmb} \chi ^{\bmc}{}_{\bmc} + \tfrac{1}{6} \kappa  \rho  h_{\bma\bmb} \chi ^{\bmc}{}_{\bmc} \\
&- E_{\bmc\bmd} h_{\bma\bmb} \chi ^{\bmc\bmd} + \epsilon _{\bmc\bmd\bme\bmf} B_{\bmb}{}^{\bme} h_{\bma}{}^{\bmf} \chi ^{\bmc\bmd} + \epsilon _{\bmd\bmf\bma} u_{\bmb} B_{\bme}{}^{\bmf}  \chi ^{\bmd\bme}.
\end{split}
\end{equation}
Similarly, the symmetric part of the term
$F^{\ast}_{\bma\bmb\mathbf{0}}$ in equation  (\ref{ReducedBianchiDual}) gives
the evolution equations for $B_{\bma\bmb}$ ---namely,
\begin{equation}
\label{EvlEqMagWeyl}
\begin{split}
 u^{\bmd} \nabla_{\bmd}B_{\bma\bmb} + D^{\bmf}E_{(\bmb}{}^{\bmd}  \epsilon _{\bma)\bmd\bmf} &=- \tfrac{1}{2} a^{\bmd} E_{\bmb}{}^{\bmf} \epsilon _{\bma\bmd\bmf} - \tfrac{1}{2} a^{\bmd} E_{\bma}{}^{\bmf} \epsilon _{\bmb\bmd\bmf}\\
 &- a^{\bmd} u_{\bmb} B_{\bma\bmd} - a^{\bmd} u_{\bma} B_{\bmb\bmd} - \tfrac{1}{2} \kappa  \Phi _{\bma\bmb} - \tfrac{1}{4} \kappa  a^{\bmd} \epsilon _{\bmb\bmd\bmf} \Pi _{\bma}{}^{\bmf}\\
 &- \tfrac{1}{4} \kappa  a^{\bmd} \epsilon _{\bma\bmd\bmf} \Pi _{\bmb}{}^{\bmf} + \tfrac{1}{2} B_{\bmb\bmd} \chi _{\bma}{}^{\bmd} + \tfrac{1}{2} B_{\bma\bmd} \chi _{\bmb}{}^{\bmd} + B_{\bmb\bmd} \chi^{\bmd}{}_{\bma}\\
 &+ B_{\bma\bmd} \chi ^{\bmd}{}_{\bmb} -2 B_{\bma\bmb} \chi ^{\bmd}{}_{\bmd} - \tfrac{1}{2} E_{\bmf}{}^{\bmc} \epsilon _{\bmb\bmd\bmc} u_{\bma} \chi ^{\bmd\bmf}\\
 &-\tfrac{1}{2} E_{\bmf}{}^{\bmc} \epsilon _{\bma\bmd\bmc} u_{\bmb} \chi^{\bmd\bmf} - B_{\bmd\bmf} h_{\bma\bmb} \chi^{\bmd\bmf}.
\end{split}
\end{equation}

We note that equations \eqref{EvolEqElWeyl} and \eqref{EvlEqMagWeyl} are on the same form as the one given in \cite{Friedrich1998} and constitutes a  symmetric hyperbolic system of equations. We refer once again to \cite{Alho2010} for an explicit discussion.

\begin{remark}
{\em The standard approach to show that equations \eqref{EvolEqElWeyl} and \eqref{EvlEqMagWeyl} constitute a symmetric hyperbolic system ignores the tracefreeness of the fields $E_{\bma\bmb}$ and $B_{\bma\bmb}$ and list 12 components in a vector. Thus, \emph{a posteriori} it is necessary to show that the fields are tracefree if they were so initially. This is discussed in Section \ref{section7}.}
\end{remark}

\subsection{Summary}
We summarise the results of the long computations of this section by the following proposition:
\begin{proposition}
The evolution equations for the matter fields as expressed by $\Pi_{\bma\bmb}$ is given by \eqref{EvolMatter1} and satisfy the constraint \eqref{EvolMatter2}. Furthermore, the evolution equations for the geometric fields $e_{\bma}{}^{\mu}$,
$\Gamma_{\bma}{}^{\bmb}{}_{\bmc}$, $E_{\bma\bmb}$, $B_{\bma\bmb}$ and the auxiliary fields $\Psi_{\bma\bmb}$, $\Phi_{\bma\bmb}$ are given, respectively,  by
\begin{subequations}
\begin{eqnarray}
&&\Sigma_{\mathbf{0}}{}^\bme{}_{\bmb} = 0, \label{EvlEqFrames}\\
&&\Delta^{\bmd}{}_{\bmc\bma\bmzero} = 0, \label{EvlEqConnection}\\
&&J_{\bmi\bma} = 0,\\
&&F_{\bma\bmb\bmzero} = 0, \label{EvlEqWeyl1}\\
&&F^{\ast}_{(\bma\bmb)\bmzero}= 0, \label{EvlEqWeyl2}\\
&&\nabla^{\bmb}N_{\bmc\bma\bmb} = 0, \label{EvlEqMatter1}\\
&&\nabla^{\bmb}N^{\ast}_{(\bmc\bma)\bmb} = 0.\label{EvlEqMatter2}
\end{eqnarray}
\end{subequations}
The above evolution equations constitutes a symmetric hyperbolic system. The remaining equations from
\eqref{DivergenceFreeCond}-\eqref{FriedrichEquation} are considered
constraint equations.
\end{proposition}

\section{Propagation equations}
\label{section6}
In order to complete our analysis of the evolution system, we need to
show that the equations that have been discarded in the process of
hyperbolic reduction (i.e. the \emph{constraints}) propagate. In this
section we will, therefore, construct a subsidiary system for the
zero-quantities $\Sigma^{\bme}{}_{\bma\bmb}$,
$\Delta^{\bmd}{}_{\mathbf{cab}}$ and $F_{\mathbf{abc}}$. The task is
then to show that either the Lie derivative of the constraints vanish,
or that it may be written in terms of zero-quantities. A key
observation in this strategy is the fact that several of the
zero-quantities can be regarded as differential forms with respect to a certain
subset of their indices ---thus, \emph{Cartan's identity} can be readily be used to
compute the Lie derivative in a very convenient way.
\begin{remark}
\em
In what follows one should be careful when evaluating the covariant derivative of a tensor fields in frame coordinates. The following order should be employed: first, evaluate the tonsorial expression for the derivative of the tensor, then write the expression in a frame basis, and lastly do any contractions if necessary --- e.g contracting with the four velocity.
\end{remark}
 \subsection{Propagation of Divergence-free condition}
 \label{subsectionTProp}
 
 The divergence free condition gives rise to two equations: on the one hand, equation \eqref{2ndEuler} is an evolution equation for $\rho$ and the other hand equation \eqref{1stEuler} a constraint for $\bm\Pi$. As such, the latter needs to be shown to hold on the whole space time if satisfied on an initial hypersurface.
 
 \smallskip
 We first define the zero quantity,
 \[
 Q_{\bmb} \equiv \nabla^{\bma}\Pi_{\bma\bmb} + a_{\bmb}\rho - u_{\bmb}\Pi_{\bma\bmc}\chi^{\bma\bmc}.
 \]
 As is obvious from the above, $Q_{\bmb}=0$ must hold for the Einstein equations to be satisfied.  By contracting with $\bmu$ it is readily shown that $Q_{\bm0} = Q_{\bmb} u^{\bmb} = 0.$ Thus it is sufficient to only consider $Q_{\bmi}$ in what follows. A simple calculation shows that,
 \[
 \begin{split}
 2\nabla_{[\bmd}Q_{\bmb]} &= 2\nabla^{\bma}Z_{\bma\bmd\bmb} + 2\nabla_{[\bmd}\left(a_{\bmb]}\rho\right)+ 2R_{\bme}{}^{\bma}{}_{[\bmd | \bma |} \Pi_{\bmb]}{}^{\bme} - 2\nabla_{[\bmd}\left(u_{\bmb]}\Pi_{\bma\bmc}\chi^{\bma\bmc}\right) + 4\Sigma_{[\bmd}{}^{\bmc}{}_{|\bma|}\nabla_{\bmc}\Pi_{\bmb]}{}^{\bma},
 \end{split}
 \]
 where we have used the commutation property of the connection followed by the definition of the Z-tensor, as well as,
 \[
 2R^{\bme}{}_{[\bma\bmd]\bma} \Pi_{\bme}{}^{\bma} = 0.
 \]
 By using equation \eqref{DivergenceZelda} and multiplying through with $u^{\bmd}$, followed by applying equations \eqref{DeltaDefinition} and \eqref{JoiCompExp}, we obtain the propagation equation,
 \begin{equation}
 \label{QPropEq}
      u^{\bma}\nabla_{\bma}Q_{\bmi} = -Q^{\bmj} \chi_{\bmi\bmj} + \Delta_{\bm0}{}{}^{\bmk\bmj}{}_{\bmk}
\Pi_{\bmi\bmj} + 2\Sigma_{\bmi}{}^{\bml}{}_{\bmj}\Pi_{\bmk}{}^{\bmj}\chi_{\bml}{}^{\bmk} + 2\Sigma_{\bmi}{}^{\bm0}{}_{\bmj}\Pi_{\bmk}{}^{\bmj}a^{\bmk}.
 \end{equation}
Note that $\rho_{\bmj}{}^{\bmk}{}_{\bmzero\bmk} = 0$ due to the divergence free property of Weyl and $T_{\bm0 \bmi} = 0$ as a consequence of the gauge. We have also made use of the evolution equation $\Sigma_{\bm0}{}^{\bmc}{}_{\bma}=0$.
 
 \subsection{Propagation equations for the torsion}
For fixed value of the index $\bme$, the torsion
$\Sigma_\bma{}^\bme{}_\bmb$ can be regarded as the components of a 2-form ---namely, one has that
\[
\bm\Sigma^{\bme} \equiv \Sigma_{[\bmb}{}^{\bme}{}_{\bmc]} \bm\omega^{\bmb} \otimes \bm\omega^{\bmc}.
\]
Using \textit{Cartan's identity} to compute its Lie derivative along the vector $\bmu$
one finds that 
\begin{equation}
\label{LieTorsion}
\mathcal{L}_{\bmu} \bm\Sigma^{\bme} = i_{\bmu} \mbox{d}\bm\Sigma^{\bme} + \mbox{d}(i_{\bmu} \bm\Sigma^{\bme}).
\end{equation}
The second term in the right-hand side of the above equation can be
seen to vanish as a consequence of the evolution equations (cf. Remark \ref{remarkSigma}) while
the first one involves the exterior derivative of the torsion which
can be manipulated using the general form of the Bianchi identity. For
clarity, these computations are done explicitly using frame index
notation.

\medskip
Following the general discussion given above consider the expression
$\nabla_{[\bmzero} \Sigma_\bma{}^\bmc{}_{\bmb]}$ which roughly
corresponds to the first term in the right-hand side of equation
\eqref{LieTorsion}. Expanding the expression one readily finds that
\begin{eqnarray*}
&& 3\nabla_{[\bmzero} \Sigma_\bma{}^\bmc{}_{\bmb]} = 
   \nabla_\bmzero \Sigma_\bma{}^\bmc{}_\bmb +
   \nabla_\bma\Sigma_\bmb{}^\bmc{}_\bmzero +
   \nabla_\bmb\Sigma_\bmzero{}^\bmc{}_\bma\\
&&\phantom{3\nabla_{[\bmzero} \Sigma_\bma{}^\bmc{}_{\bmb]} } = \nabla_\bmzero \Sigma_\bma{}^\bmc{}_\bmb - \Gamma_\bma{}^\bme{}_\bmzero \Sigma_\bmb{}^\bmc{}_\bme - \Gamma_\bmb{}^\bme{}_\bmzero \Sigma_\bma{}^\bmc{}_\bme.
\end{eqnarray*}
Now, we compute $\nabla_{[\bmzero}
\Sigma_\bma{}^\bmc{}_{\bmb]}$ in a different way using the general
expression for the first Bianchi identity (i.e. the form this identity
takes in the presence of torsion):
\[
R^{\bmd}{}_{[\bmc\bma\bmb]} = -
\nabla_{[\bma}\Sigma_\bmb{}^\bmd{}_{\bmc]} - \Sigma_{[\bma}{}^{\bme}{}_\bmb
\Sigma_{\bmc]}{}^{\bmd}{}_\bme. 
\]
Setting $\bma=0$ and making use of the zero-quantity defined in
\eqref{DeltaDefinition} to eliminate the components of the Riemann
curvature tensor one finds that 
\begin{eqnarray*}
&& 3\nabla_{[\mathbf{0}}\Sigma_\bmb{}^{\bmd}{}_{\bmc]}  = -
\Delta^{\bmd}{}_{[\bmc \bmzero\bmb]}-
   \Sigma_{[\bmzero}{}^\bme{}_\bmb\Sigma_{\bmc]}{}^\bmd{}_\bme\\
&& \phantom{3\nabla_{[\mathbf{0}}\Sigma_\bmb{}^{\bmd}{}_{\bmc]}} = -\Delta^\bmd{}_{\mathbf{0}\bmb\bmc}
\end{eqnarray*}
where we have used the fact that
\[
\rho^{\bmd}{}_{[\bmc\bma\bmb]} = 0
\]
and the evolution equations \eqref{EvlEqFrames} and \eqref{EvlEqConnection} in the last step. From the above discussion it follows the \emph{propagation equation}
\begin{equation}
 \nabla_\bmzero \Sigma_\bma{}^\bmc{}_\bmb=  \Gamma_\bma{}^\bme{}_\bmzero \Sigma_\bmb{}^\bmc{}_\bme + \Gamma_\bmb{}^\bme{}_\bmzero \Sigma_\bma{}^\bmc{}_\bme  -\Delta^\bmd{}_{\mathbf{0}\bmb\bmc}.
 \label{PropagationEq:Sigma}
\end{equation}

\begin{remark}
{\em The main structural feature of equation \eqref{PropagationEq:Sigma} is the fact that it is homogeneous in the zero-quantities $\Sigma_\bma{}^\bmc{}_\bmb$ and $\Delta^\bmd{}_{\bma\bmb\bmc}$.}
\end{remark}


\subsection{Propagation equations for the geometric curvature}

Next we turn to equation \eqref{DeltaEquation}. For this we
observe that the zero-quantity $\Delta^\bmd{}_{\bmc\bma\bmb}$ for
fixed values of $\bmd$ and $\bmc$ can be regarded as the components of
a 2-form on the indices $\bma$ and $\bmb$. Using again Cartan's
identity one finds that
\[
\mathcal{L}_{\mathbf{u}} \bm\Delta^{\bmd}{}_{\bmc} = i_{\bmu} \mbox{d} \bm\Delta^{\bmd}{}_{\bmc} + \mbox{d}\left(i_{\bmu} \bm\Delta^{\bmd}{}_{\bmc} \right).
\]
Now, the last term in the right-hand side vanishes due to the
evolution equation for the connection coefficients (see Remark \ref{Remark:VanishingDelta}),
while the first term takes the form
\[
i_{\mathbf{u}} \mbox{d} \mathbf{\Delta}^{\bmd}{}_{\bmc} = \nabla_{[\mathbf{0}} \Delta^{\bmd}{}_{|\bmc|\bma\bmb]}\bm\omega^{\bma} \otimes \bm\omega^{\bmb}.
\]
As in the case of the torsion,  the strategy is to rewrite this
expression in terms of zero-quantities only. For convenience in the following calculations we set
\begin{equation}
\label{DefSTensor}
S_{\bma\bmb}{}^{\bmc\bmd} \equiv \delta_{\bma}{}^{\bmc} \delta_{\bmb}{}^{\bmd} + \delta_{\bma}{}^{\bmd}
\delta_{\bmb}{}^{\bmc} -\eta_{\bma\bmb}\eta^{\bmc\bmd}.
\end{equation}
From equations \eqref{2ndBianchiId} and \eqref{rhoDef} it readily follows that 
\begin{equation*}
\begin{split}
\nabla_{[\bma} \Delta^{\bmd}{}_{|\bme|\bmb\bmc]} = \nabla_{[\bma}C^{\bmd}{}_{|\bme|\bmb\bmc]} - S^{\bmd\bmf}{}_{\bme[\bmb}\nabla_{\bma} \hat{L}_{\bmc]\bmf} - \Sigma^{\bmf}{}_{[\bma\bmb}R^{\bmd}_{|\bme|\bmc]\bmf}.
\end{split}
\end{equation*}
To simplify the calculations, we multiply by $\epsilon_{\bml}{}^{\bma\bmb\bmc}$. The first term yields
\begin{equation}
\label{calc0}
\epsilon_{\bml}{}^{\bma\bmb\bmc} \nabla_{[\bma}C^{\bmd}{}_{|\bme|\bmb\bmc]} 
 = -2\nabla_{\bma}{C^{\ast}}\indices{^{\bma}_{\bml}^{\bmd}_{\bme}}. 
\end{equation}
while the second term gives 
\begin{equation}
\label{calc1}
\begin{split}
\epsilon_{\bml}{}^{\bma\bmb\bmc} S_{\bme[\bmb}{}^{\bmd\bmf}\nabla_{\bma} \hat{L}_{\bmc]\bmf} = -\epsilon_{\bml}{}^{\bmd\bma\bmc} \nabla_{\bma} \hat{L}_{\bmc\bme} + \epsilon_{\bml\bme}{}^{\bma\bmc} \eta^{\bmd\bmf}\nabla_{\bma} \hat{L}_{\bmc \bmf}
\end{split}
\end{equation}
where we have made use of the definition \eqref{DefSTensor} and the symmetry of the Schouten tensor.
Now, from equations \eqref{FriedrichDefinition}, \eqref{FriedrichDef} and \eqref{FriedrichEquation} we readily
obtain that  
\begin{equation*}
\epsilon_{\bml}{}^{\bmd\bmb\bmc} \nabla_{[\bmb}\hat{L}_{\bmc]\bma} = - \epsilon_{\bml}{}^{\bmd\bmb\bmc} F_{\bma\bmb\bmc} + \epsilon_{\bml}{}^{\bmd\bmb\bmc}\nabla_{\bmm}C^{\bmm}{}_{\bma\bmb\bmc}.
\end{equation*}
Plugging this result back into \eqref{calc1}, we obtain
\begin{equation}
\label{calc2}
\begin{split}
\epsilon_{\bml}{}^{\bma \bmb\bmc} S_{\bme[\bmb}{}^{\bmd\bmf}\nabla_{\bma} \hat{L}_{\bmc]\bmf} &=  \epsilon_{\bml}{}^{\bmd\bma\bmc} F_{\bme\bma\bmc} - \eta^{\bmd\bmf}\epsilon_{\bml\bme}{}^{\bma \bmc} F_{\bmf\bma\bmc}\\
&+ \eta^{\bmd\bmf} 2\nabla_{\bma}C^{\ast}{}^{\bma}{}_{\bmf\bml\bme} - 2\nabla_{\bma}C^{\ast}{}^{\bmb\bmm\bma}{}_{\bme\bml}{}^{\bmd}.
\end{split}
\end{equation}
Putting the result for calculation \eqref{calc2} and \eqref{calc0} together, we obtain
\begin{equation*}
\begin{split}
\epsilon_{\bml}{}^{\bma\bmb\bmc} \nabla_{\bma}\rho^{\bmd}{}_{\bme\bmb\bmc} &=  \epsilon_{\bml}{}^{\bmd\bma\bmc} F_{\bme\bma\bmc} - \eta^{\bmd\bmf}\epsilon_{\bml\bme}{}^{\bma\bmc} F_{\bmf\bma\bmc},
\end{split}
\end{equation*}
where we made use of equation \eqref{1stBianchiId}. Multiplying by $\epsilon^{\bml}{}_{\bmm\bmn\bmp}$, we recover the equation in its original form. Thus, the right-hand side of the propagation equation for $\Delta^{\bmd}{}_{\bmc[\bma\bmb]}$ is given by

\begin{equation}
\label{PropEqDelta1}
\nabla_{[\bm0} \Delta^{\bmd}{}_{|\bme|\bmb\bmc]} = - \eta_{\bme[\bm0} F^{\bmd}{}_{\bmb\bmc]} + \eta^{\bmd}{}_{[\bm0} F_{|\bme|\bmb\bmc]}. 
\end{equation}
But we also have that,
\[
\nabla_{[\bm0} \Delta^{\bmd}{}_{|\bme|\bmb\bmc]} = \nabla_{\bm0}\Delta^{\bmd}{}_{\bme\bmb\bmc} - \Gamma_{\bmc}{}^{\bmf}{}_{\bm0}\Delta^{\bmd}{}_{\bme\bmf\bmb} - \Gamma_{\bmb}{}^{\bmf}{}_{\bm0}\Delta^{\bmd}{}_{\bme\bmc\bmf}.
\]
Plugging the above result back into equation \eqref{PropEqDelta1}, we obtain the final propagation equation
\begin{equation}
\begin{split}
\label{PropEqDelta}
  \nabla_{\bm0}\Delta^{\bmd}{}_{\bme\bmb\bmc} =  - \eta_{\bme[\bm0} F^{\bmd}{}_{\bmb\bmc]} + \eta^{\bmd}{}_{[\bm0} F_{|\bme|\bmb\bmc]}
  +\Gamma_{\bmc}{}^{\bmf}{}_{\bm0}\Delta^{\bmd}{}_{\bme\bmf\bmb} + \Gamma_{\bmb}{}^{\bmf}{}_{\bm0}\Delta^{\bmd}{}_{\bme\bmc\bmf}.
  \end{split}
\end{equation}
\begin{remark}
{\em As in the case of the propagation equation for the torsion the main conclusion of the previous discussion is that the propagation equation for the zero-quantitity $\Delta^\bmd{}_{\bma\bmb\bmc}$ is homogeneous on zero-quantities.}
\end{remark}

\subsection{Propagation of the N-tensor}
It is also necessary to show that $N_{\bma\bmb\bmc}$ ---see equation \eqref{NDef}--- propagates. The strategy will be different than what has been employed in the above discussions; rather, we will follow the strategy employed for the propagation of the Friedrich tensor in \cite{Friedrich1996}.

\smallskip
In the subsequent discussion we shall make use of the observation that, 
\[
 \nabla^{\bmb}N'^{\ast}{}_{\bmc\bma\bmb} = 0,\qquad N'{}_{\bmc\bm0\bmb} = 0,\qquad N'{}^{\ast}{}_{\bmc\bm0\bmb} = 0.
\]
respectively are equivalent to the evolution equations for $\Psi_{\bma\bmb}$, $\Phi_{\bma\bmb}$ and $\Pi_{\bma\bmb}$ and the constraint equation as given in \eqref{EvolMatter2}.
Furthermore, we define the fields,
\[
\xi_{\bma\bmb} \equiv N'^{\ast}{}_{\bma\bmb\bm0}, \qquad \lambda_{\bma\bmb} \equiv N'{}_{\bm0\bma\bmb}.
\]
By decomposing $N^{\ast}{}_{\bma\bmb\bmc}$ in terms of the fields $\lambda_{\bma\bmb}$ and $\xi_{\bma\bmb}$, we have
\begin{equation}
\label{NDualDecomp}
N^{\ast}{}_{\bmc\bma\bmb} = \xi_{\bmc\bma}u_{\bmb} - \xi_{\bmc\bmb}u_{\bma} + 3\lambda_{\bmd\bme}u_{[\bmb}\epsilon_{\bma]}{}^{\bmd\bme}u_{\bmc}
\end{equation}
Using the symmetry relation
\[
N_{[\bma\bmb\bmc]} = 0,
\]
we obtain the expression
\[
\lambda_{\bma\bmb} = N'{}_{\bmb\bma\bm0} - N'{}_{\bma\bmb\bm0}.
\]
But from the evolution equation for $\Pi_{\bma\bmb}$, we have that $N'{}_{\bmb\bma\bm0}=0$, thus
\[
\lambda_{\bma\bmb} = 0.
\]
Applying the above result, and the divergence in equation \eqref{NDualDecomp} we obtain,
\[
u_{\bmb}\nabla^{\bmb}\xi_{\bmc\bmd} = \xi_{\bmc\bma}a^{\bma}u_{\bmd} - \xi_{\bmc\bmd}\chi + \xi_{\bmc\bmb}\chi^{\bmb}{}_{\bmd}.
\]
To obtain the above we have used the evolution equation for $\Phi_{\bma\bmb}$ and multiplied through with the projector $h_{\bmd}{}^{\bma}$ to get rid of a divergence. This is permitted as the field $\xi_{\bma\bmb}$ is spatial. Thus, we have established the following lemma:
\begin{lemma}
If the constraint $\xi_{\bma\bmb} = 0$ --- equivalently equation \eqref{EvolMatter2} --- holds initialy, and under the assumption that the evolution equations \eqref{EvolMatter1} and \eqref{EvlEqMatter2}
holds everywhere on $\mathcal{M}$, then the relation,
\[
Z_{\bmc\bma\bmb} = 2\nabla_{[\bma}\Pi_{\bmb]\bmc},
\]
also holds everywhere on $\mathcal{M}.$
\end{lemma}
\subsection{Propagation equations for the Bianchi identity}
Lastly, we need to show propagation of the Bianchi identity, equation
\eqref{FriedrichEquation}. Again the strategy is to use the decomposition of the Friedrich tensor and its dual and use the divergence to obtain propagation equations for the constraints. 

\smallskip
First, we shall express the divergence of $F_{\bma\bmb\bmc}$ in terms of known zero quantities. We have,
\[
\begin{split}
2\nabla^{\bmb} F_{\bmb\bmc\bmd} &= -2R^{\bml}{}_{\bma}{}^{\bmb\bma} C_{\bml\bmb\bmc\bmd} -2R^{\bml}{}_{\bmb}{}^{\bmb\bma} C_{\bma\bml\bmc\bmd} -2R^{\bml}{}_{\bmc}{}^{\bmb\bma} C_{\bma\bmb\bml\bmd}\\
&-2R^{\bml}{}_{\bmd}{}^{\bmb\bma} C_{\bma\bmb\bmc\bml} + 2R^{\bml}{}_{\bmd}{}^{\bmb}{}_{\bmc}\hat{L}_{\bml\bmb} + 2R^{\bml}{}_{\bmb}{}^{\bmb}{}_{\bmc}\hat{L}_{\bmd\bml}\\
&- R^{\bml}{}_{\bmc}{}^{\bmb}{}_{\bmd}\hat{L}_{\bml\bmb} - R^{\bml}{}_{\bmb}{}^{\bmb}{}_{\bmd}\hat{L}_{\bmc\bml} + 4\Sigma_{\bma}{}^{\bml}{}_{\bmb}\nabla_{\bml}C^{\bma\bmb}{}_{\bmc\bmd} + \nabla_{\bmd}\nabla^{\bmb}\hat{L}_{\bmc\bmb} - \nabla_{\bmc}\nabla^{\bmb}\hat{L}_{\bmd\bmb},
\end{split}
\]
where we have used the antisymmetry property of the Weyl tensor about the indices $\bma$ and $\bmb$ and the commutator as defined in
\[
\nabla_{[\bmb}\nabla_{\bma]}\omega_{\bmc} = -R^{\bme}{}_{\mathbf{c a b}}\omega_{\bme} + \Sigma_{\bmb}{}^{\bmd}{}_{\bma}\nabla_{\bmd}\omega_{\bmc}.
\]
Substituting from  the definition \eqref{DeltaDefinition} we obtain after a lengthy and somewhat messy calculation that 
\begin{equation}
\label{PropFriedrichFinal}
\begin{split}
\nabla^{\bmb} F_{\bmb\bmc\bmd} &= -\Delta^{\bml}{}_{\bma}{}^{\bmb\bma} C_{\bml\bmb\bmc\bmd} -\Delta^{\bml}{}_{\bmb}{}^{\bmb\bma} C_{\bma\bml\bmc\bmd} -\Delta^{\bml}{}_{\bmc}{}^{\bmb\bma} C_{\bma\bmb\bml\bmd}
-\Delta^{\bml}{}_{\bmd}{}^{\bmb\bma} C_{\bma \bmb\bmc\bml}\\
&+ \Delta^{\bml}{}_{\bmd}{}^{\bmb}{}_{\bmc}\hat{L}_{\bml\bmb} + \Delta^{\bml}{}_{\bmb}{}^{\bmb}{}_{\bmc}\hat{L}_{\bmd\bml}
- \Delta^{\bml}{}_{\bmc}{}^{\bmb}{}_{\bmd}\hat{L}_{\bml\bmb} - \Delta^{\bml}{}_{\bmb}{}^{\bmb}{}_{\bmd}\hat{L}_{\bmc\bml} + \nabla_{[\bmd}Q_{\bmc]}\\
&+ 4\Sigma_{\bma}{}^{\bml}{}_{\bmb}\nabla_{\bml}C^{\bma\bmb}{}_{\bmc\bmd} - 2\Sigma_{\bmb}{}^{\bml}{}_{\bmc}\nabla_{\bml}S_{\bmd}{}^{\bmb} + 2\Sigma_{\bmb}{}^{\bml}{}_{\bmc}\nabla_{\bml}S_{\bmc}{}^{\bmb} + \frac{1}{3}\Sigma^{\bme}{}_{\bmd\bmc}\nabla_{\bme}T,
\end{split}
\end{equation}
where $q_{\bma}$ is the zero quantity defined in section \ref{subsectionTProp}.

Following the strategy outlined in \cite{Friedrich1996}, we define the fields,
\[
p_{\bma} \equiv F'_{\bm0\bma\bm0}, \qquad q_{\bma} \equiv F^{\ast\prime}_{\bm0\bma\bm0},
\]
which encodes the information of the constraint equations of $F_{\bma\bmb\bmc}$ and $F^{\ast}_{\bma\bmb\bmc}$, respectively. Thus, the aim is to find evolution equations for $p_{\bma}$ and $q_{\bma}$.

In terms of the above fields the decomposition \eqref{ReducedBianchi} takes the form,
\begin{equation}
\label{DecompFriedrichConstraints}
F_{\bmb\bmc\bmd} = 2u_{\bmb}p_{[\bmc}u_{\bmd]} + h_{\bmb[\bmd}p_{\bmc]} - \frac{1}{2}u_{\bmb}q_{\bme}\epsilon_{\bmc\bmd}{}^{\bme},
\end{equation}
where we have used,
\[
F'_{\bm0\bmb\bmd} = q_{\bme}\epsilon_{\bmb\bmd}{}^{\bme}, \qquad F'_{\bmb\bmc\bmd} = h_{\bmb[\bmd}p_{\bmc]} + \frac{1}{2}u_{\bmb}q_{\bme}\epsilon_{\bmc\bmd}{}^{\bme}.
\]
To obtain the above, we used the evolution equations ---i.e. that $F_{\bmb\bmc\bm0} = 0$ and $F^{\ast}{}_{(\bmb\bmc)\bm0}=0$ as well as the identity,
\[
F_{[\bma\bmb]\bmc} = -\frac{1}{2}F_{\bmc\bma\bmb},
\]
which is a direct result of the symmetry properties of $F_{\bma\bmb\bmc}.$
By taking the divergence of the first index of \eqref{DecompFriedrichConstraints} and equating with $u^{\bmd}h_{\bma}{}^{\bmc}$ times \eqref{PropFriedrichFinal}, we obtain a propagation equation for the $p_{\bma}$ field, namely 
\begin{equation}
\begin{split}
2u^{\bmb}\nabla_{\bmb}p_{\bma}&= 2u_{\bma}a^{\bmc}p_{\bmc} - \epsilon_{\bma}{}^{\bmc\bmd}a_{\bmc}q_{\bmd}+\chi_{\bma}{}^{\bmc}p_{\bmc}
+3\chi p_{\bma}+2\Delta^{\bml}{}_{\bma}{}^{\bmb\bmm} C_{\bml\bmb\bmc\bm0}h_{\bma}{}^{\bmc}\\ &+2\Delta^{\bml}{}_{\bmb}{}^{\bmb \bmm} C_{\bmm\bml\bmc\bm0} h_{\bma}{}^{\bmc}+2\Delta^{\bml}{}_{\bmc}{}^{\bmb\bmm} C_{\bmm\bmb\bml\bm0}h_{\bma}{}^{\bmc}+2\Delta^{\bml}{}_{\bm0}{}^{\bmb\bmm} C_{\bmm\bmb\bmc\bml}h_{\bma}{}^{\bmc}\\
&+2 \Delta^{\bml}{}_{\bmc}{}^{\bmb}{}_{\bm0}\hat{L}_{\bml \bmb}h_{\bma}{}^{\bmc} +2 \Delta^{\bml}{}_{\bmb}{}^{\bmb}{}_{\bm0}\hat{L}_{\bma \bml}
-2 \nabla_{[\bm0}Q_{\bmc]}h_{\bma}{}^{\bmd}\\
&+ 4u^{\bmd}h_{\bma}{}^{\bmc}\Sigma_{\bma}{}^{\bml}{}_{\bmb}\nabla_{\bml}C^{\bma\bmb}{}_{\bmc\bmd} - 2u^{\bmd}h_{\bma}{}^{\bmc}\Sigma_{\bmb}{}^{\bml}{}_{\bmc}\nabla_{\bml}S_{\bmd}{}^{\bmb} + 2u^{\bmd}h_{\bma}{}^{\bmc}\Sigma_{\bmb}{}^{\bml}{}_{\bmc}\nabla_{\bml}S_{\bmc}{}^{\bmb}.
\end{split}
\label{Equation:E}
\end{equation}
In the above, we have used that $\Sigma_{\bm0}{}^{\bmd}{}_{\bmb} = 0$ everywhere on $\mathcal{M}$. Applying the same procedure to \eqref{ReducedBianchiDual} we obtain the propagation equations for the $q_{\bma}$ field,

\begin{equation}
\begin{split}
u^{\bmb}\nabla_{\bmb}q_{\bma} - \epsilon_{\bma}{}^{\bmc\bmd}\mathcal{D}_{\bmc}p_{\bmd} &= u_{\bma}a^{\bmc}q_{\bmc} - \chi q_{\bma}+2p_{\bmc}\epsilon_{\bma\bmb}a^{\bmb}\\
&+\Delta^{\bml}{}_{\bma}{}^{\bmb\bmm} C^{\ast}_{\bml\bmb\bmc\bm0}h_{\bma}{}^{\bmc} +\Delta^{\bml}{}_{\bmb}{}^{\bmb \bmm} C^{\ast}_{\bmm\bml\bmc\bm0} h_{\bma}{}^{\bmc}+\Delta^{\bml}{}_{\bmp}{}^{\bmb\bmm} C_{\bmm\bmb\bml\bmn}\epsilon_{\bma}{}^{\bmp\bmn}\\
&+\Delta^{\bml}{}_{\bmn}{}^{\bmb\bmm} C_{\bmm\bmb\bmp\bml}\epsilon_{\bma}{}^{\bmp\bmn} + \Delta^{\bml}{}_{\bmp}{}^{\bmb}{}_{\bmn}\hat{L}_{\bml \bmb}\epsilon_{\bma}{}^{\bmp\bmn} + \Delta^{\bml}{}_{\bmb}{}^{\bmb}{}_{\bmn}\hat{L}_{\bmp \bml}\epsilon_{\bma}{}^{\bmp\bmn}\\
&+\nabla_{\bmn}Q_{\bmp}\epsilon_{\bma}{}^{\bmn\bmp} + 4u^{\bmm}\Sigma_{\bma}{}^{\bml}{}_{\bmb}\nabla_{\bml}C^{\ast}{}^{\bma\bmb}{}_{\bmm\bma} + 2\Sigma_{\bmb}{}^{\bml}{}_{\bmn}\nabla_{\bml}S_{\bmp}{}^{\bmb}\epsilon_{\bma}{}^{\bmn\bmp}.
\end{split}
\label{Equation:B}
\end{equation}

\begin{remark}
{\em Again, the main observation to be extracted from the previous analysis is that equations \eqref{Equation:E} and \eqref{Equation:B} are homogeneous in the various zero-quantities. Moroever, their form is analogous to that of the evolution equations \eqref{EvolEqElWeyl} and \eqref{EvlEqMagWeyl}. Thus, it can be verified they imply a symmetric hyperbolic system.
Note also that equation \eqref{Equation:E} is different in the principle part compared to equation \eqref{Equation:B}. This is due to thhat the evolution equation $F_{\bma\bmb\bm0} = 0$ is not symmetrized. It is also understood in equation \eqref{Equation:E} that one apply equation \eqref{QPropEq} to eliminate the time derivative of $Q_{\bma}$.}
\end{remark}

\subsection{Main theorem}

The homogeneity of the propagation equations for the various zero-quantities implies, from the uniqueness of symmetric hyperbolic systems that if the zero-quantities vanish on some initial hypersurface $\mathcal{S}_\star$ then they will also vanish at later times. We summarise the analysis of the previous subsections in the following statement:

\begin{theorem}
A solution 
\[
(e_{\bma}{}^{\mu}, \Gamma_{\bmi}{}^{\bmj}{}_{\bmk},
\Gamma_{\bm0}{}^{\bm0}{}_{\bmk}, \Gamma_{\bmi}{}^{\bm0}{}_{\bmj},
E_{\bma\bmb}, B_{\bma\bmb}, \Psi_{\bma\bmb}, \Phi_{\bma\bmb},
\Pi_{\bma\bmb}, \rho)
\]
to the system of evolution equations given, respectively, by equations 
\eqref{EvolutionEquationCoefficientsFrame}, \eqref{EvolutionEquarionCoefficientsConnection},
\eqref{evolConnection2}, \eqref{evolConnection3},
\eqref{EvolEqElWeyl}, \eqref{EvlEqMagWeyl}, \eqref{EvolMatter1},
\eqref{PsiEvolutionEq},\eqref{EvolEqPhi} and
\eqref{LieRho} with initial data satisfying the conditions
\[
\Sigma_\bma{}^\bmb{}_\bmc=0, \qquad \Delta^\bmd{}_{\bma\bmb\bmc}=0, \qquad F_{\bma\bmb\bmc}=0,
\]
on an initial hypersurface $\mathcal{S}_\star$ implies a solution to the Einstein-matter frame equations \eqref{DivergenceFreeCond}-\eqref{FriedrichEquation}.
\end{theorem}

\begin{remark}
{\em As a consequence of Lemma \ref{Lemma:SolutionFrameSolutionEinstein} it follows that the solution the Einsteinmatter frame equations implies, in turn, a solution to the standard Einstein-matter field equations \eqref{EFE}.}
\end{remark}



\section{Matter models}
\label{section7}
We will in the following exemplify the previous discussion with a
number of particular matter models. We shall also note that although the equations given in the following resembles those found in \cite{Friedrich1998},  the treatment of the propagation of constraints for dust or perfect fluid was not treated therein. We fill this gap in this paper.

\subsection{Dust}
The simplest case is of course that of \textit{dust}. In this case
$\Pi_{\bma\bmb} = 0$ and the expression for the energy-momentum
tensor, equation \eqref{EnergyMom}, reduces to
\[
T_{\bma\bmb} = \rho u_{\bma}u_{\bmb}.
\]
Furthermore, as there are no internal interactions, each dust particle follows a geodesic ---i.e the following hold
\[
a_{\bma} = 0, \qquad \Gamma_{\bm0}{}^{\bmc}{}_{\bmb} = 0.
\]
Consequently, equation \eqref{LieRho} reduces to
\begin{equation}
u^{\bma}\nabla_{\bma}\rho = -\rho \chi \label{LieRhoDust}
\end{equation}
and  equation \eqref{EvolutionEquarionCoefficientsConnection} takes the form
\begin{equation}
\begin{split}
\label{EvolutionEquarionCoefficientsConnectionDust}
\partial_{\mathbf{0}} \Gamma_{\bmi}{}^{\bma}{}_{\bmb} &= - \Gamma_{\bmd}{}^{\bmc}{}_{\bmb}\Gamma_{\bmi}{}^{\bmd}{}_{\mathbf{0}} - C^{\bma}{}_{\bmb\bmi\bmzero}.
\end{split}
\end{equation}
Also, we have that $Z_{\bma\bmb\bmc} = 0$. Thus, the discussion of the
$Z$-tensor and its evolution equations are irrelevant ---i.e. there is no need for the construction of an auxiliary field. The evolution equations for the Weyl tensor reduce to
\begin{equation}
\label{EvolEqElWeylDust}
\begin{split}
u^{\bmc} \nabla_{\bmc}E_{\bma\bmb} -\epsilon _{\bma\bme\bmf} D^{\bmf}B_{\bmb}{}^{\bme}  &=  - \tfrac{1}{2} \kappa  \rho  \chi _{\bmb\bma} - E_{\bma\bmc} \chi _{\bmb}{}^{\bmc} + 2 E_{\bmb\bmc} \chi ^{\bmc}{}_{\bma} - 2 E_{\bma\bmb} \chi ^{\bmc}{}_{\bmc} + \tfrac{1}{6} \kappa  \rho  h_{\bma\bmb} \chi ^{\bmc}{}_{\bmc} \\
&- E_{\bmc\bmd} h_{\bma\bmb} \chi ^{\bmc\bmd} + \epsilon _{\bmc\bmd\bme\bmf} B_{\bmb}{}^{\bme} h_{\bma}{}^{\bmf} \chi ^{\bmc\bmd} + \epsilon _{\bmd\bmf\bma} u_{\bmb} B_{\bme}{}^{\bmf}  \chi ^{\bmd\bme} ,
\end{split}
\end{equation}
and,
\begin{equation}
\label{EvlEqMagWeylDust}
\begin{split}
 u^{\bmd} \nabla_{\bmd}B_{\bma\bmb} + D^{\bmf}E_{(\bmb}{}^{\bmd}  \epsilon _{\bma)\bmd\bmf} &= \tfrac{1}{2} B_{\bmb\bmd} \chi _{\bma}{}^{\bmd} + \tfrac{1}{2} B_{\bma\bmd} \chi _{\bmb}{}^{\bmd} + B_{\bmb\bmd} \chi^{\bmd}{}_{\bma} + B_{\bma\bmd} \chi ^{\bmd}{}_{\bmb} -2 B_{\bma\bmb} \chi ^{\bmd}{}_{\bmd}\\
 & - \tfrac{1}{2} E_{\bmf}{}^{\bmc} \epsilon _{\bmb\bmd\bmc} u_{\bma} \chi ^{\bmd\bmf} -\tfrac{1}{2} E_{\bmf}{}^{\bmc} \epsilon _{\bma\bmd\bmc} u_{\bmb} \chi^{\bmd\bmf} - B_{\bmd\bmf} h_{\bma\bmb} \chi^{\bmd\bmf}.
\end{split}
\end{equation}
Thus, equations \eqref{EvolutionEquationCoefficientsFrame},
\eqref{EvolutionEquarionCoefficientsConnectionDust},
\eqref{EvolEqElWeylDust}, \eqref{EvlEqMagWeylDust} and
\eqref{LieRhoDust} provide the symmetric hyperbolic evolution
equations for the fields $e_{\bma}^{\mu}$,
$\Gamma_{\bmi}{}^{\bma}{}_{\bmb}$, $E_{\bma\bmb}$, $B_{\bma\bmb}$ and
$\rho$, respectively.

\subsection{Perfect fluid}
Before we discuss the details of a perfect fluid, we shall briefly
review some important quantities in relativistic thermodynamics.

\smallskip
Given a material with $N$ different particle species, $n_{A}$ denote
the number density of a particular species, where $A =
\{1,2,...,N\}$. Furthermore, we denote by $s$ the entropy density. The
energy density of the system is a function of these quantities ---i.e. we have
\begin{equation}
\rho = f\left(s, n_{1}, n_{2},...,n_{N}\right).\label{EqOfState}
\end{equation}
The function $f$ is called the \textit{equation of state} of the system.
Finally, the \textit{first law of Thermodynamics} is given by,
\begin{equation}
d\rho = Tds + \mu^Adn_A,\label{1stLawofThermoDyn} 
\end{equation}
where,
\[
T \equiv \left(\dfrac{\partial \rho}{\partial s}\right)_{n_A}, \qquad \mu_{A} \equiv \left(\dfrac{\partial \rho}{\partial n_A}\right)_{s},
\]
denotes the temperature and chemical potential, respectively.  In what
follows we shall consider a simple perfect fluid --- i.e a fluid of
only one type of particles ($A = 1$) and with an energy momentum
tensor with
\begin{equation}
\Pi_{\bma\bmb} = p h_{\bma\bmb}.\label{PiDefPF} 
\end{equation}
Consequently, we have
\[
\Pi_{\{\bma\bmb\}} = p u_{\bma}u_{\bmb}, \qquad \Pi = 3p,
\]
where $p$ denote the pressure and is defined by
\begin{equation}
p \equiv n \mu - \rho.\label{DefPreassure}
\end{equation}
Throughout we shall assume an equation of state of the form given by
\eqref{EqOfState} with $A = 1$ and the law of particle conservation
---i.e.
\begin{equation}
u^{\bma}\nabla_{\bma}n = - n\chi.\label{ParticleConservation}
\end{equation}
With these assumptions, equations \eqref{DivStressTraceFree} and
\eqref{LieRhoTraceFree} reduce to the well known Einstein-Euler
equations, given by
\begin{subequations}
\begin{eqnarray}
&& u_{\bmb}u^{\bma}\nabla_{\bma}p + \nabla_{\bmb}p = - \left(\rho + p\right) a_{\bmb}, \label{1stEuler}\\
&& u^{\bma}\nabla_{\bma} \rho = -\left(\rho +p\right)\chi.\label{2ndEuler}
\end{eqnarray}
\end{subequations} 
It follows from equations \eqref{DefPreassure}, \eqref{2ndEuler},
\eqref{ParticleConservation} and \eqref{1stLawofThermoDyn} that the
fluid is adiabatic ---i.e we have
\begin{equation}
u^{\bma}\nabla_{\bma}s = 0.\label{EvolEqEntropy}
\end{equation}
From the above discussion it follows that equation \eqref{evolConnection2} takes the form
\begin{equation}
\begin{split}
3\partial_{\bm0} \Gamma_{\bm0}{}^{\bm0}{}_{\bmi} -
\eta^{\bmj\bmk}\partial_{\bmi}\Gamma_{\bmj}{}^{\bm0}{}_{\bmk} &= - 2a_{\bmi}\chi + a^{\bmj}\chi_{\bmi\bmj} - \partial^{\bmj}\Gamma_{\bmj}{}^{\bm0}{}_{\bmi} +\Gamma_{\bmj}{}^{\bmk}{}_{\bmi}\chi_{\bmk}{}^{\bmj} -\Gamma_{\bmj}{}^{\bmj}{}_{\bmk}\chi_{\bmi}{}^{\bmk}\\
&- \frac{1}{\rho}\left(R_{\bm0}{}_{\bmi} p + 2p a_{\bmi} \chi\right) - \Gamma_{\bmj}{}^{\bm0}{}_{\bm0}\chi^{\bmj}{}_{\bmi}- \Gamma_{\bm0}{}^{\bm0}{}_{\bmi}\chi\\
&+ \Gamma_{\bm0}{}^{\bmj}{}_{\bmi}\Gamma_{\bm0}{}^{\bm0}{}_{\bmj} + \eta^{\bmj\bmk}\Gamma_{\bmi}{}^{\bml}{}_{\bmj}\Gamma_{\bml}{}^{\bm0}{}_{\bmk} + \eta^{\bmj\bmk}\Gamma_{\bmi}{}^{\bml}{}_{\bmk}\Gamma_{\bmj}{}^{\bm0}{}_{\bml}. \label{evolConnectionPF1}.
\end{split}
\end{equation}
Similarly, equation
\eqref{evolConnection3} takes the form
\begin{equation}
\label{evolConnectionPF2}
\begin{split}
\partial_ {\bm0}\Gamma_{\bmi}{}^{\bm0}{}_{\bmj}- \partial_{\bmi}\Gamma_{\bm0}{}^{\bm0}{}_{\bmj} &= \frac{2}{\rho}R_{[\bmj}{}^{\bmk} p h_{\bmi]\bmk} + 2 \chi_{\bmk[\bmi}\chi_{\bmj]}{}^{\bmk} + \frac{4}{\rho}\left(\mu a_{[\bmj} n_{\bmi]} + n T a_{[\bmj}s_{\bmi]}\right) - R^{\bm0}{}_{\bmi\bm0\bmj}\\
&-\Gamma_\bmk{}^{\bm0}{}_{\bmi}\Gamma_{\bmj}{}^\bmk{}_{\bm0} -\Gamma_\bmj{}^\bmk{}_\bmi\Gamma_{\bm0}{}^{\bm0}{}_{\bmk} - \Gamma_{\bmi}{}^{\bm0}{}_{\bm0}a_{\bmj} + \Gamma_{\bmj}{}^{\bm0}{}_{\bm0}a_{\bmi}\\
&+ \Gamma_{\bm0}{}^{\bmk}{}_{\bmi}\Gamma_{\bmk}{}^{\bm0}{}_{\bmj}+ \Gamma_{\bm0}{}^{\bmk}{}_{\bmj}\Gamma_{\bmi}{}^{\bm0}{}_{\bmk} + \Gamma_{\bm0}{}^{\bmk}{}_{\bmi}\Gamma_{\bm0}{}^{\bm0}{}_{\bmk} - \Gamma_{\bm0}{}^{\bmk}{}_{\bmj}\Gamma_{\bm0}{}^{\bm0}{}_{\bmk}.,
\end{split}
\end{equation}
where we have defined
\[
s_{\bma} \equiv \nabla_{\bma}s, \qquad n_{\bma} \equiv \nabla_{\bma}n.
\]
The corresponding evolution equations are obtained by using equations \eqref{EvolEqEntropy} and \eqref{ParticleConservation}, 
\begin{subequations}
\begin{eqnarray}
&& u^{\bma}\nabla_{\bma}s_{\bmb} = -s_{\bma}\left(\chi_{\bmb}{}^{\bma} + a^{\bma}u_{\bmb}\right),\label{EvolDivn}\\
&& u^{\bma}\nabla_{\bma}n_{\bmb} = - n_{\bmb}\chi - n\nabla_{\bmb}\chi -n_{\bma}\left(\chi_{\bmb}{}^{\bma} + a^{\bma}u_{\bmb}\right)\label{EvolDivs}.
\end{eqnarray}
\end{subequations}
Now, writing equations \eqref{EvolMatter1} and \eqref{EvolMatter2} in terms of the above definitions we obtain
\begin{subequations}
\begin{eqnarray}
&&\Psi_{\bma\bmb} = \chi_{\bmb\bma} - h_{\bma\bmb}\left(\rho + p\right)\left(1 - \nu^2\right)\chi + h_{\bma\bmb}\mu n \chi,\label{PsiSub}\\
&&\Phi_{\bma\bmb} = -\epsilon_{\bma}{}^{\bmd}{}_{\bmb}\left(\rho + p\right)a_{\bmd} - \epsilon_{\bma\bmc\bmd}\chi^{\bmc\bmd}p u_{\bmb}\label{PhiSub},
\end{eqnarray}
\end{subequations}
where we have used equation \eqref{DefPreassure} and the definition of $\mu$ to obtain,
\begin{equation}
u^{\bma}\nabla_{\bma}p = \left(\rho + p\right)\left(1 - \nu^2\right)\chi - \mu n \chi.\label{EvolEqP}
\end{equation}

Finally, the evolution equations for $E_{\bma\bmb}$ and $B_{\bma\bmb}$
are obtained by substituting equations \eqref{PsiSub}, \eqref{PhiSub}
and \eqref{PiDefPF} into the equations \eqref{EvolEqElWeyl} and
\eqref{EvlEqMagWeyl}
\begin{equation}
\label{EvolutionEPF}
\begin{split}
u^{\bmc} \nabla _{\bmc}E_{\bma\bmb} - \epsilon _{\bma\bmc\bmd} D^{\bmd}B_{\bmb}{}^{\bmc} &= -2a^{\bmc} u_{(\bma}E_{\bmb)\bmc} +  a^{\bmc} \epsilon_{\bmb\bmc\bmd} B_{\bma}{}^{\bmd} - 2\kappa  a_{(\bmb} u_{\bma)} p  +  \kappa  p \chi _{[\bma\bmb]} +  \tfrac{1}{2} \kappa  \chi _{\bmb\bma}\\
&-\tfrac{1}{2} \kappa  \rho  \chi _{\bmb\bma} + E_{\bma\bmc} \chi _{\bmb}{}^{\bmc} + 2 E_{\bmb\bmc} \chi ^{\bmc}{}_{\bma} - 2 E_{\bma\bmb} \chi ^{\bmc}{}_{\bmc} - \tfrac{1}{6} \kappa  h_{\bma\bmb} \chi ^{\bmc}{}_{\bmc}\\
&+  \tfrac{1}{6} \kappa  \rho  h_{\bma\bmb} \chi ^{\bmc}{}_{\bmc} + \epsilon_{\bma\bmc\bme} u_{\bmb} B_{\bmd}{}^{\bme} \chi ^{\bmc\bmd} - E_{\bmc\bmd} h_{\bma\bmb} \chi ^{\bmc\bmd},
\end{split}
\end{equation}
\begin{equation}
\label{EvolutionBPF}
\begin{split}
u^{\bmd} \nabla _{\bmd}B_{\bma\bmb} - D^{\bmf}E_{(\bma}{}^{\bmd} \epsilon
_{\bmb)\bmd\bmf} &= - \tfrac{1}{2} a^{\bmd} E_{\bmb}{}^{\bmf} \epsilon _{\bma\bmd\bmf} - \tfrac{1}{2} a^{\bmd} E_{\bma}{}^{\bmf} \epsilon _{\bmb\bmd\bmf} - a^{\bmd} u_{\bmb} B_{\bma\bmd} - a^{\bmd} u_{\bma} B_{\bmb\bmd}+ \tfrac{1}{2} B_{\bmb\bmd} \chi _{\bma}{}^{\bmd}\\
&+ \tfrac{1}{2} B_{\bma\bmd} \chi _{\bmb}{}^{\bmd}+ B_{\bmb\bmd} \chi ^{\bmd}{}_{\bma} + B_{\bma\bmd} \chi^{\bmd}{}_{\bmb} -2 B_{\bma\bmb} \chi ^{\bmd}{}_{\bmd} - \tfrac{1}{2} E_{\bmf}{}^{\bmc} \epsilon _{\bmb\bmd\bmc} u_{\bma} \chi ^{\bmd\bmf}\\
&- \tfrac{1}{2} E_{\bmf}{}^{\bmc} \epsilon _{\bma\bmd\bmc} u_{\bmb} \chi ^{\bmd\bmf} - B_{\bmd\bmf} h_{\bma\bmb} \chi ^{\bmd\bmf} + \tfrac{1}{4} \kappa  \epsilon _{\bmb\bmd\bmf} u_{\bma} p \chi ^{\bmd\bmf} + \tfrac{1}{4} \kappa  \epsilon _{\bma\bmd\bmf} u_{\bmb} p \chi ^{\bmd\bmf} .
\end{split}
\end{equation}
Equations \eqref{EvolutionEquationCoefficientsFrame},
\eqref{EvolutionEquarionCoefficientsConnection},
\eqref{evolConnectionPF1}, \eqref{evolConnectionPF2},
\eqref{EvolutionEPF}, \eqref{EvolutionBPF}, \eqref{EvolEqP},
\eqref{2ndEuler}, \eqref{EvolDivs}, \eqref{EvolDivn},
\eqref{EvolEqEntropy} and \eqref{ParticleConservation} provide the
symmetric hyperbolic system for the fields $(e_{\bma}^{\mu}$,
$\Gamma_{\bmi}{}^{\bmj}{}_{\bmk}$, $\Gamma_{\bm0}{}^{\bm0}{}_{\bmk}$,
$\Gamma_{\bmi}{}^{\bm0}{}_{\bmj}$, $E_{\bma\bmb}$, $B_{\bma,\bmb}$,
$p$, $\rho$, $s_{\bmi}$,$n_{\bmi}$,$s$,$n)$ respectively.

\subsection{Elastic matter}
The following discussion follows the treatment of relativistic elasticity found in \cite{Beig2003a}. 

\medskip
The energy density of the elastic system is given by
\begin{equation}
\label{LagrangianElastic}
\rho = n\epsilon,
\end{equation}
where $\epsilon$ is the \textit{stored energy function} of the
system. It can be shown\footnote{See Appendix
\ref{Appendix:Elasticity}.} 
that the elastic energy-momentum tensor in frame coordinates can be
put on the form of equation \eqref{EnergyMom} with an energy density
as given by \eqref{LagrangianElastic} and
\begin{equation}
\label{DefPiElastic}
\Pi_{\bma\bmb} \equiv 2\rho \eta_{\bma\bmb}+2n \tau_{\mathbf{AB}} \Lambda^{\bmA}{}_{\bma}\Lambda^{\bmB}{}_{\bmb},
\end{equation}
where $\tau_{\bmA\bmB}$ denotes the relativistic \textit{Cauchy stress tensor} and is defined by
\[
\tau_{\bmA\bmB} \equiv \dfrac{\partial \epsilon}{\partial h^{\bmA\bmB}},
\]
where $h^{\bmA\bmB}$ is the \textit{strain tensor} defined by
\[
h^{\bmA\bmB} \equiv \Lambda^{\mathbf{A}}{}_{\mathbf{a}}\Lambda^{\mathbf{B}}{}_{\mathbf{b}} \eta^{\bma\bmb}.
\]
The particle density number $n$ is also defined in terms of these
fields --- i.e. we have that 
\[
n^2 \equiv \frac{1}{6}\det(h^{\bmA\bmB}).
\]
Thus, the components $\Lambda^{\bmA}{}_{\bma}$ is the fundamental material field of
the theory (see the Appendix for more details). We shall, however, not write explicit equations for these
fields, but rather use the formalism described in the main part of the
paper. Hence, the information regarding $\Lambda_{\bma}{}^{\bmA}$ is
encoded in the tensor $\Pi_{\bma\bmb}$ by equation
\eqref{DefPiElastic}. Consequently, the symmetric hyperbolic system
for the fields  $(e_{\bma}^{\mu}$, $\Gamma_{\bmi}{}^{\bmj}{}_{\bmk}$,
$\Gamma_{\bm0}{}^{\bm0}{}_{\bmk}$, $\Gamma_{\bmi}{}^{\bm0}{}_{\bmj}$,
$E_{\bma\bmb}$, $B_{\bma,\bmb}$, $\Psi_{\bma\bmb}$, $\Phi_{\bma\bmb}$,
$\Pi_{\bma\bmb}$, $\rho)$ are respectively given by equations
\eqref{EvolutionEquationCoefficientsFrame},
\eqref{EvolutionEquarionCoefficientsConnection},
\eqref{evolConnection2}, \eqref{evolConnection3},
\eqref{EvolEqElWeyl},\eqref{EvlEqMagWeyl}, \eqref{EvolMatter1},
\eqref{PsiEvolutionEq},\eqref{EvolEqPhi} and
\eqref{LieRho}. Equations \eqref{DivStress} and \eqref{EvolMatter2}
are considered constraint equations.

 \section{Concluding remarks}
 \label{section8}
As stressed previously, we have developed first order symmetric hyperbolic evolution equations for a wide range of matter models which solves the Einstein equations. Our formalism should thus be applicable to the development of a theory of Neutron stars as an relativistic elastic system. In this case one would proceed as with the perfect fluid case: one need to write  the tensor $\bm\Pi$ in terms of its trace and trace-free part and provide equations for $n$ and $\epsilon$ to close the system. The latter is likely obtained from thermodynamical considerations. In addition it is necessary to provide an equation for $\bm\sigma$ in terms of lower order terms.

The treatment given in this paper is sufficiently general that showing
symmetric hyperbolicity for a given matter model coupled to the Einstein equations, is reduced to the
simple task of showing that the system admits an energy momentum tensor on the form \eqref{EnergyMom} satisfying \eqref{PiSym} and \eqref{PiSpatial}. It is understood that an equation for $\rho$ and $\bm\sigma$ is provided.

\appendix
\section{The frame components of elastic energy-momentum tensor}
\label{Appendix:Elasticity}

In the following let $\mathcal{B}$ be a 3-dimentional manifold
representing the ensemble of particles making up the elastic body. The
key object in relativistic elasticity is a map 
\[
\phi: \mathcal{M} \rightarrow \mathcal{B},
\]
 such that if $\overline{x}=(x^{\mu})$ and $\overline{X}=(X^{M})$ are, respectively,
 coordinates on the spacetime and body manifold we then have
\[
\phi^M(x^{\mu}) = X^M.
\]
As the manifolds $\mathcal{M}$ and $\mathcal{B}$ have, respectively,
dimension 4 and 3, the map $\phi$ is non-injective
(one-to-one). In the following it will be assumed that the inverse
image $\phi^{-1}(\overline{X})$ of a point on $\mathcal{B}$ with
coordinates $\overline{X}=(X^M)$ is a timelike curve on
$\mathcal{M}$. We denote the tangent vector to the
curve $\gamma:\mathbb{R}\rightarrow \mathcal{M}$ with 
$\gamma \equiv \phi^{-1}(\overline{X})$ by $\bmu$. If we assume $\gamma$
to be parametrised by its proper time, then we have that
\[
\bmg(\bmu, \bmu) =-1.
\]
The curve $\gamma$ describes the worldline of the particle of the
point on $\mathcal{B}$ with coordinates
$\overline{X}$.

\medskip
The map $\phi$ represents the \textit{configuration} of the elastic
body. This means that $\phi$ associates to each spacetime event a
material particle. In other words, $\phi$ relates the physical state
of a material body with the notion of an event in spacetime. The
deformation of the elastic body is represented by the \textit{deformation
  gradient}, defined by in terms of the coordinates at $\mathcal{M}$
and $\mathcal{B}$ by 
\[
 \phi^M{}_\mu = \partial_\mu \phi^M.
\]
For a fixed value of the coordinate indices on the body manifold, the
components $ \phi^M{}_\mu$ give rise to a covector field $\phi^M{}_a$
on $\mathcal{M}$ satisfying the condition
\[
\phi^M{}_a u^a = 0.
\]
We introduce the \textit{strain} of the material by applying the
\textit{push-forward} to the inverse metric tensor $g^{\mu \nu}$ on
$\mathcal{M}$. Its the components are given by
\[
h^{MN} \equiv \phi^M{}_{\mu} \phi^N{}_{\nu} g^{\mu \nu}.
\]
The body manifold is equipped with a volume form $V_{ABC}$ which
allows us to construct a scalar function $n$ interpreted as the
\textit{particle density number} of the material via the relation
\[
n^2 = \frac{1}{3!} \det (h^{MN}).
\]
This interpretation of $n$ is found reasonable by the observation that
the equation for particle conservation hold ---that is, one has that
\[
\nabla_{\mu}\left(n u^{\mu}\right) = 0.
\]

In order to formulate a frame version of the energy momentum tensor of a relativistic elastic material, we
begin by consider a frame $\{\bmE_{\bmA}\}$ on  $\mathcal{B}$ with
associated coframe $\{ \bm\Omega^{\bmB} \}$. As we have not introduced a
metric on $\mathcal{B}$, we do not assume any orthonormality condition
on the frame and coframe.

\medskip
The map $\phi$, defines the pullback $\phi^{\ast}$ which can be used to
pull-back the coframe to $\mathcal{M}$. More precisely, one has that
\[
\bm\Lambda^{\bmB}  \equiv  \phi^{\ast} \bm\Omega^{\bmB}, \qquad \Lambda^{\bmB}{}_{\bma} = \langle\bm\Lambda^{\bmB},\bme_{\bma}\rangle.
\]
As the map $\phi$ is surjective and has maximal rank, the set of
covectors $\{ \bm\Lambda^{\bmB}\}$ is linearly independent. The fields
$\{\Lambda^{\bmB}{}_{\bma}\}$ will be used, in the sequel, to describe the
configuration of the material body. The coefficients $\Lambda^\bmA{}_\bma$ are \emph{orthogonal} with
respect to $u^a$ ---that is
\[
\Lambda^{\mathbf{A}}{}_{\bma} u^\bma = 0.
\]

We denote the determinant of the frame field as $e$. It is related to
the determinant of the metric tensor by $e = \sqrt{-g}$. Furthermore, we have
\begin{subequations}
\begin{eqnarray}
&&\delta e = \omega_{\bma}{}^{\mu} \delta e_{\bma}^{\mu}, \label{deltae}\\
&&\delta \Lambda^{\bmA}{}_{\bma} = \Lambda^{\bmA}{}_{\mu} \delta e_{\bma}{}^{\mu}. \label{deltaLambda}
\end{eqnarray}
\end{subequations}
In the above $\delta$ is understood as an infinitesimal variation. More precicely, for a function $f$ we have,
\[
\delta f \equiv \dfrac{\partial f}{\partial x^{\mu}}\delta x^{\mu}.
\]
Equation \eqref{deltae} can be obtained by using \textit{Jacobi's formula} for a square matrix given by,
\[
\begin{split}
\dfrac{\partial \det(\bmA)}{\partial A_{\mu\nu}} =det(\bmA) (\bmA^{-1})_{\mu\nu},
\end{split}
\]
and recalling that $\omega^{\bma}{}_{\mu} = (e^{-1})^{\bma}{}_{\mu}$.
Equation \eqref{deltaLambda} follows form observing that 
\[
\Lambda^{\bmA}{}_{\bma} =\Lambda^{\bmA}{}_{\mu} e_{\bma}{}^{\mu}, \qquad \dfrac{\partial e_{\bma}{}^{\nu}}{\partial  e_{\bmc}{}^{\mu}} =  \delta^{\bmc}{}_{\bma}\delta^{\nu}{}_{\mu}.
\]
\medskip
In terms of the above fields we construct a Lagrangian on the form $L = L\left(\Lambda^{\bmA}{}_{\bmb},e_{\bma}{}^{\mu}\right)
 $. The action thus reads
\[
S = \int \mathcal{L}\left(\Lambda^{\bmA}{}_{\bmb},e_{\bma}{}^{\mu}\right) d^4x,
\]
where we have defined the \textit{Lagrangian density}
\[
\mathcal{L}\left(\Lambda^{\bmA}{}_{\bmb},e_{\bma}{}^{\mu}\right) \equiv e L\left(\Lambda^{\bmA}{}_{\bmb},e_{\bma}{}^{\mu}\right).
\]
The variation of the action yields
\[
\begin{split}
\delta S &= \int \left(\dfrac{\partial e}{\partial e_{\bma}{}^{\mu}}\delta e_{\bma}{}^{\mu} L +  e \dfrac{\partial L}{\partial e_{\bma}{}^{\mu}}\delta e_{\bma}{}^{\mu} + e \dfrac{\partial L}{\partial \Lambda^{\bmA}{}_{\bmb}}\delta \Lambda^{\bmA}{}_{\bmb}\right)d^4x\\
&= \int \left(\omega^{\bma}{}_{\mu}L + \dfrac{\partial L}{\partial e_{\bma}{}^{\mu}} + \dfrac{\partial L}{\partial \Lambda^{\bmA}{}_{\bma}} \Lambda^{\bmA}{}_{\mu}\right) e \delta e_{\bma}{}^{\mu}d^4x\\
&= \int \mathcal{T}^{\bma}{}_{\mu} e \delta e_{\bma}{}^{\mu}d^4x,
\end{split}
\]
where we have made use of equations \eqref{deltae} and \eqref{deltaLambda} and defined
\[
 \mathcal{T}^{\bma}{}_{\mu} \equiv \omega^{\bma}{}_{\mu}L + \dfrac{\partial L}{\partial e_{\bma}{}^{\mu}} + \dfrac{\partial L}{\partial \Lambda^{\bmA}{}_{\bma}} \Lambda^{\bmA}{}_{\mu}.
\]
By multiplying with $e_{\bmc}{}^{\mu} \eta^{\bma\bmc}$, and applying the chain rule to the second term, we obtain
\begin{equation}
\label{EMDefinition}
 \mathcal{T}^{\bma\bmb} = \eta^{\bma\bmb}L + 2\dfrac{\partial L}{\partial \Lambda^{\bmA}{}_{\bma}}  \Lambda^{\bmA}{}_{\mathbf{c}}\eta^{\bmb\bmc}.
\end{equation}
Assuming that the Lagrangian may be written on the form (see
\cite{Wernig-Pichler2006} for details) 
\[
L = \rho = n \epsilon,
\]
we find that
\[
\dfrac{\partial L}{\partial \Lambda^{\bmA}{}_{\bmb}}  = n \dfrac{\partial \epsilon}{\partial h^{\bmA\bmB}} \dfrac{\partial h^{\bmA\bmB}}{\partial \Lambda^{\bmA}{}_{\bmb}}+  \epsilon \dfrac{\partial n}{\partial \Lambda^{\bmA}{}_{\bmb}}
\]
with
\[
\dfrac{\partial h^{\bmA\bmB}}{\partial \Lambda^{\bmD}{}_{\bma}}=
2\eta^{\bma\bmc}\Lambda^{(\bmA}{}_{\bmc}\delta^{\bmB)}{}_{\bmD},
\qquad \dfrac{\partial n}{\partial \Lambda^{\bmD}{}_{\bma}} = n h_{\mathbf{AB}}\eta^{\bma\bmc}\Lambda^{(\bmA}{}_{\bmc}\delta^{\bmB)}{}_{\bmd}.
\]
Substituting the above expressions back into equation
\eqref{EMDefinition} we find an expression for the components of the
energy-momentum tensor of the form 
\begin{equation}
\label{EMDefinitionFinal}
T^{\bma\bmb}= n \epsilon \eta^{\bma\bmb} + \Pi^{\bma\bmb}
\end{equation}
where, in the following, $\Pi^{\bma\bmb}$ will be known as the
components of the \emph{Cauchy stress tensor} and is given by
\begin{equation}
\Pi_{\bma\bmb} \equiv 2n \tau_{\mathbf{AB}} \Lambda^{\bmA}{}_{\bma}\Lambda^{\bmB}{}_{\bmb} + \epsilon n h_{\mathbf{AB}} \Lambda^{\bmA}{}_{\bma}\Lambda^{\bmB}{}_{\bmb}.\label{PiDefOne}
\end{equation}
We further make the reasonable assumption that
\[
h_{\bma\bmb} = h_{\bmA\bmB} \Lambda^{\bmA}{}_{\bma}\Lambda^{\bmB}{}_{\bmb},
\]
where $h_{\bma\bmb}$ as usual denotes the frame components of the
projector metric. To show that this is reasonable, we note the
following: the equation holds identically both under multiplication of $u^{\bma}$ and $\eta^{\bmc\bma}
\Lambda^{\bmC}{}_{\bma}$  --- in the latter case, one has to
invoke the definition of $h_{\bmA\bmB}$ on the right hand side of the equation to show this.
Secondly, on a spatial hypersurface $\mathcal{S} \in \mathcal{M}$ the map
$\phi$ is a diffeomorphism which implies that the object
$h_{\bma\bmb}$ defined on $\mathcal{S}$ is physically equivalent to the
corresponding object defined on $\mathcal{B}$ via $\phi$. Using this
assumption in \eqref{PiDefOne} we obtain the desired form of the
energy momentum tensor. Namely, one has that 
\begin{equation}
\label{EMDefinitionFinal2}
T_{\bma\bmb}= \rho u_{\bma}u_{\bmb} + \Pi_{\bma\bmb},
\end{equation}
with
\begin{equation}
\Pi_{\bma\bmb} \equiv 2\rho \eta_{\bma\bmb}+2n \tau_{\mathbf{AB}} \Lambda^{\bmA}{}_{\bma}\Lambda^{\bmB}{}_{\bmb}. \label{PiDefTwo}
\end{equation}


\end{document}